\documentclass[fleqn,usenatbib]{mnras}

\usepackage{newtxtext,newtxmath}

\usepackage[T1]{fontenc}
\usepackage{ae,aecompl}
\usepackage{color}

\usepackage[figure,figure*]{hypcap}
\usepackage{tabularx}
\usepackage{tabu}
\usepackage{wrapfig,layout,paralist,color}

\usepackage{graphicx}	
\usepackage{amsmath}	
\usepackage{amssymb}	

\usepackage{etoolbox}
\makeatletter
\patchcmd\@combinedblfloats{\box\@outputbox}{\unvbox\@outputbox}{}{%
   \errmessage{\noexpand\@combinedblfloats could not be patched}%
}%
 \makeatother

\title{Updating the MACHO fraction of the Milky Way dark halo with improved mass models}

\author[Calcino et al.]{Josh Calcino,$^{1}$\thanks{Contact e-mail: \href{mailto:j.calcino@uq.edu.au}{j.calcino@uq.edu.au}}
Juan Garc\'{i}a-Bellido,$^{2}$
Tamara M. Davis$^{1}$ \\ 
$^{1}$School of Mathematics and Physics, The University of Queensland, QLD 4072, Australia\\
$^{2}$Instituto de F\'{i}sica Te\'{o}rica UAM-CSIC, Universidad Aut\'{o}noma de Madrid, Cantoblanco, 28049 Madrid, Spain \\
}

\date{Last updated \today; in original form 2013 September 5}

\pubyear{2018}

\begin{document}
\label{firstpage}
\pagerange{\pageref{firstpage}--\pageref{lastpage}}
\maketitle

\begin{abstract}
Recent interest in primordial black holes as a possible dark matter candidate has motivated the reanalysis of previous methods for constraining massive astrophysical compact objects in the Milky Way halo and beyond. In order to derive these constraints, a model for the dark matter distribution around the Milky Way must be used. 
Previous microlensing searches have assumed a semi-isothermal density sphere for this task. We show this model is no longer consistent with data from the Milky Way rotation curve, and test two replacement models, namely NFW and power-law. The power-law model is the most flexible as it can break spherical symmetry, and best fits the data. 
Thus, we recommend the power-law model as a replacement, although it still lacks the flexibility to fully encapsulate all possible shapes of the Milky Way halo. We then use the power-law model to rederive some previous microlensing constraints in the literature, while propagating the primary halo-shape uncertainties through to our final constraints.

Our analysis reveals that the microlensing constraints towards the Large Magellanic Cloud weaken somewhat for MACHO masses around $10\, M_\odot$ when this uncertainty is taken into account, but the constraints tighten at lower masses. Exploring some of the simplifying assumptions of previous constraints we also study the effect of wide mass distributions of compact halo objects, as well as the effect of spatial clustering on microlensing constraints. We find that both effects induce a shift in the constraints towards smaller masses, and can effectively remove the microlensing constraints from $M \sim 1-10 M_\odot$ for certain MACHO populations.
\end{abstract}




\section{Introduction}
Interest in Primordial Black Holes (PBHs) as a dark matter (DM) candidate has been growing since the gravitational wave discoveries from LIGO of black hole (BH) binary systems in the mass range $\sim$ 10$M_\odot$ \citep{LigoFirst2016, LigoSecond2016}. PBHs would act as MACHOs, and would therefore be detectable from their gravitational influence as they pass between an observer and a distant source object, and from their dynamical effects on their surroundings. Multiple microlensing surveys have been, and are still, being conducted to discover such objects \citep{alcock2000, eros2007, ogle2011, ogle2015}. Collectively these surveys constrain MACHOs in the mass range $10^{-7} \lesssim M/M_\odot \lesssim 30$ primarily by frequent observations of the Large and Small Magellanic Clouds (LMC and SMC). 

Constraints on the higher mass end of the spectrum also exist. The dynamical effects from MACHOs will disrupt wide binary stars in the galactic halo, putting an upper limit on the fraction of MACHOs as DM in the halo \citep{chanam2004,yoo2004,quinn2009,monroy2014}. Similarly, dynamical heating of dwarf galaxies will be evident if MACHOs constitute a large fraction of DM, but this dynamical heating is not observed \citep{eri2016,brandt2016,koush2017}. However, the existence of an intermediate mass black hole at the centre of the dwarf galaxy would weaken any MACHO constraints \citep{Li2016}. If PBHs make up a substantial fraction of DM, then accretion onto these PBHs from the interstellar medium should be observable with radio and X-ray telescopes, but again this is not observed \citep{gaggero2016,inoue2017}. Accretion of matter onto PBHs in the early universe will also affect light from the CMB and the reionisation history of the universe, constraining the existence of PBH with $M \gtrsim 10\, M_\odot$ \citep{chen2016, aloni2017, ali2017}. See \citep{Poulin2017} for a detailed analysis of gas accretion onto PBH before recombination and their effect on the CMB constraints. 

Collectively these results strongly disfavour the hypo\-thesis that PBHs make up a dominant component of DM, although there is still room for PBHs to make up a small fraction. However, there are some details and assumptions in obtaining these constraints that are overlooked and could significantly change the conclusion. In particular, we wish to focus on the constraints obtained by microlensing.

The first detail is that the constraints are generally obtained by assuming a delta-function mass distribution. Inflationary models that produce PBHs do so with an extended mass function \citep{carr2010,green2014,clesse2015,garcia2017}, which will inevitably change the constraints compared to when a delta-function mass distribution is assumed. \cite{carr2016} argue that when an extended mass distribution is taken into account, PBHs in the intermediate-mass window ($1-10^3 M_\odot$) can still make the entirety of DM. However their analysis is criticised by \cite{green2016}, who argues that the microlensing constraints from EROS-2 \citep{eros2007} and the dynamical heating constraints by \cite{brandt2016} still strongly disfavour PBHs with an extended mass function.

The second detail generally overlooked is that PBHs may not be uniformly distributed through space, but may be contained in spatially small clusters~\citep{clesse2015,clesse2017,Ezquiaga2017, juan2017}. If the cluster size is sufficiently small (i.e., much smaller than the Einstein radius of the entire cluster), then the cluster itself may act as a single lens, despite being made of many smaller components~\citep{7hints2017}. In this case, it is the entire cluster mass, and not the individual PBH mass, that is constrained by microlensing searches.

A final detail that is overlooked is that the MACHO microlensing constraints are obtained by assuming a ``standard" halo \citep[e.g.\ see][]{alcock1996, alcock2000, eros2007,ogle2011}. This standard halo was primarily used in the early days of MACHO surveys since there were no reliable Milky Way Rotation Curve (MWRC) data out to the distance of the LMC, and therefore no accurate way to probe the underlying mass distribution of the MW. As pointed out by the MACHO collaboration in their first year results paper \citep{alcock1996}, accurate measurements of the Galactic mass out to large radii can be combined with microlensing observations to improve the constraints on the MACHO fraction in the MW dark halo. However this has not yet been addressed, and it is still common for MACHO fractions to be derived using the standard halo model.

Since this standard halo model has been assumed, and no longer appears to be an accurate fit to MWRC data \citep{sofue2009}, there is an intrinsic uncertainty in the microlensing constraints that is not properly accounted for. This issue has been recognised by \cite{hawkins2015}, who argue that low mass halo models can be found that are consistent with the \cite{alcock2000} optical depth, but \cite{hawkins2015} does not actually fit any MWRC data before coming to this conclusion. More recently, \cite{green2017} quantified this uncertainty, but also did so without fitting any MWRC data. Instead they made upper and lower bounds on the likely form of the rotation curve and rederived the microlensing constraints with an extended mass function from there. 

Large scale surveys now exist which enable the recovery of the MWRC out to large radii, often much further out than the LMC or SMC \citep{xue2008,sofue2009,sofue2012,sofue2013,bhat2014,sofue2015,huang2016}. 
It now seems appropriate to test the standard model against the recent data, and derive updated microlensing constraints if the model is no longer consistent with observational data.

In this paper, we test the consistency of the standard dark halo model used to derive microlensing constraints with recent data of the MWRC. We also test two other dark halo models, in search for a suitable replacement for the standard model, which we find is no longer consistent with MWRC data. We then update the EROS-2 LMC bright stars sample and MACHO Collaboration 5.7 year results to best reflect current knowledge of the MWRC, and the shape of the MW dark halo. Constraints from other microlensing surveys also exist, most notably those produced by the Optical Gravitational Lensing Experiment \citep[OGLE][]{oglesmc2011, ogle2011}. The MACHO constraints obtained by OGLE are consistent with those obtained by EROS-2, although the results from the MACHO Collaboration are in slight tension with the results from these two surveys. We discuss this further in section \ref{sec:res}.

On top of testing other halo models, we also explore how different populations of PBHs can affect the microlensing constraints. We investigate a spatially {\em uniform} distribution with a broad lognormal mass distribution, as well as a spatially {\em clustered} distribution of PBH, with $N_{\rm cl}=10-50$ PBH per cluster as a typical number. We find that in both cases the constraints are weakened.
In the case of clustering, we find that the constraints shift to lower mass by a fixed amount that only depends on the number of PBH in the cluster and the dispersion of the lognormal.

The structure of the paper is as follows: in section \ref{sec:mw} we introduce the mass models used for fitting the MWRC, we then follow on in section \ref{sec:mwud} with the MWRC dataset we use for fitting, and a discussion on sources of systematic error that are present in our analysis, along with the results from our fits. In section \ref{sec:micro} we introduce key concepts in microlensing theory. We then present our results for the constraints on MACHOs in section \ref{sec:res} before providing a summary in section \ref{sec:sum}.

\section{Milky Way Mass Models} \label{sec:mw}
Obtaining accurate information on the MWRC curve outside the Solar radius ($\sim 8.5$ kpc) has been a challenging task for astronomers. Inside the Solar radii, techniques employing simple trigonometric relationships allow one to recover an accurate picture of the MWRC (e.g. tangent-point method). Indeed an accurate rotation curve inside the Solar radius has existed for some time \citep{fich1989}.

Compiling the rotation curve outside of the Solar radius requires an accurate understanding of the 3 dimensional motion of the Sun and the distance to all sources. The distance to a source star is typically determined using the distance modulus in the case of standard candle stars, though the error from this can be quite large ($\sim$ 5-20\%). Current generation surveys overcome this by using large samples of stars. For example, the compilation by \cite{huang2016} uses roughly 22,000 stars to compile the MWRC out to a distance of 100 kpc. Datasets now exist that allow a reasonably accurate recovery of the MWRC out to large galactocentric distances \citep{sofue2012, sofue2015, bhat2014,huang2016}, allowing us to test the standard model used to derived microlensing constraints.

\subsection{Milky Way Rotation Curve}
In this paper we parametrise the MWRC as a three component model with the bulge, the disk, and the dark halo. Each of the components is characterised by some mass distribution such that the circular rotation velocity of each component is given by
\begin{equation}
V_a(R) = \sqrt{\frac{GM_a(R)}{R}},
\end{equation}
where the velocity $V_a(R)$, and $M_a(R)$ is the mass contained within $R$ of component $a$. Thus, for our three components, the total mass within $R$ is given by
\begin{equation}
M_t(R) = M_b(R) + M_d(R) + M_h(R).
\end{equation}
This will then give the circular rotation curve $V(R)$ at galactocentric radius $R$ by
\begin{equation}\label{eq:vrot}
V_t(R)^2 = \frac{GM_t(R)}{R} = V_b(R)^2 + V_d(R)^2 + V_h(R)^2,
\end{equation}
where $V_b(R)$, $V_d(R)$, and $V_h(R)$ are the circular rotational velocity components from the bulge, disk, and dark halo respectively \citep{sofue2013}. The radius $R$ lies in the galactic mid-plane ($z= 0$ pc, where $z$ is the perpendicular distance from the plane of the disk). The contribution of each of these models varies with $R$, such that the bulge component is almost negligible at large $R$ compared to the combined effect from the disk and dark halo. 
\subsection{The bulge component}
In our analysis we parametrise the bulge using an exponential sphere model \citep{sofue2017}. The volume mass density $\rho$ for this model is given by
\begin{equation}
\rho(R) = \rho_c \exp{\left(-\frac{R}{a_b}\right)},
\end{equation}
where $\rho_c$ is the central bulge density and $a_b$ is the bulge scaling radius. The mass within a sphere of some galactocentric radius $R$ is
\begin{equation}
M(R) = M_b F(x),
\end{equation}
where $M_b=8\pi a_b^3 \rho_c$ is the total bulge mass, $x = R/a_b$, and
\begin{equation}
F(x) = 1- e^{-x}(1+x+x^2/2).
\end{equation}
The function $F(x)$ can be obtained by performing a volume integral over exponential density sphere
\begin{equation}
M_b F(x) = 4\pi \rho_c \int^{x}_0 x'^2 e^{-x'} dx'.
\end{equation}
The circular rotation velocity at $R$ is finally given by
\begin{equation}
V(R) = \sqrt{\frac{GM_b}{R}F(R/a_b)}.
\end{equation}
The two free parameters are $M_b$ and $a_b$, which are set to fixed values in our analysis (see section \ref{sec:mwrctec}). Although it is well known that the bulge of the Milky Way contains a bar structure \citep{freud1998}, this has little effect on the regions of the MWRC that we are interested in. Hence the assumption of a spherically symmetric bulge is justified.

\subsection{The disk component}
In this analysis we do not split up the disk into its two known components, the thin and the thick disk. Instead it is modelled as a single disk with some scaling radius and total mass. It common to model the Milky Way disc using the `thin disk' approximation, where the height scale of the disk is neglected \citep[e.g., see][]{xue2008, sofue2012, sofue2013, huang2016}. The surface density of the disk is modelled as an exponential and is given by

\begin{equation}
\Sigma_d(R) = \Sigma_0 \exp \left( -\frac{R}{a_d} \right),
\end{equation}
where $\Sigma_0$ is the central surface density value and is given by

\begin{equation}
\Sigma_0 = \frac{M_d}{2\pi a_d^2},
\end{equation}
where $M_d$ is the total mass of the disk and $a_d$ is the disk scaling radius. An analytic solution for the rotational velocity exists
\begin{equation}
V_d(R) = \sqrt{\frac{GM_d}{a_d}}D(X),
\end{equation}
where $X=\frac{R}{a_d}$ and $D(X)$ is given by
\begin{equation}
D(X) = \frac{X}{\sqrt{2}} \left[ I_0(X/2)K_0(X/2) - I_1(X/2)K_1(X/2) \right]^{1/2},
\end{equation}
where $I_i$ and $K_i$ are the modified Bessel functions \citep{freeman1970}. We use the two parameters $M_d$ and $a_d$ for fitting this model. Not accounting for the true 3-dimensional shape of the disk has only a very small effect on the resulting MWRC fits \citep{binney2008}.

\begin{table*}
    \centering
    \caption{A summary of the models being used for fitting the MWRC in our analysis.}
    \label{tab:mw_components}
    \begin{tabular}{ccc}
        \hline
	Component & Fitted Parameters & Description \\ 
	\hline
	Bulge & $M_b$, $a_b$ &   Exponential sphere    \\ 
	Disk & $M_d$, $a_d$  &    Razor thin, exponential disk    \\ 
	Halo - semi-isothermal & $\rho_\odot$, $R_c$  &    Spherically symmetric, non-converging mass   \\ 
         - NFW &  $\rho_\odot$, $R_c$   & Spherically symmetric  \\ 
	     - power-law & $v_0$, $\beta$, $R_c$  &   Can break spherical symmetry    \\ 
    \hline \hline
    \end{tabular}
\end{table*}

\subsection{The dark halo components}
We aim to test and compare several different dark halo models by fitting the MWRC data, and then comparing the resulting expected number of microlensing events from each. In particular, we wish to test the semi-isothermal dark halo model \citep{begeman1991}, the NFW model \citep{nfw1996}, and the power-law model of \cite{evans1993, evans1994}.

\subsubsection{Semi-isothermal model}
The semi-isothermal model is the `standard' model used by many when determining constraints on the fraction of DM as MACHOs. It produces a flat rotation curve at large galactrocentric radii when the contribution from the disk component becomes negligible. The density profile is written as \citep{alcock1996}
\begin{equation}
\rho(R) = \rho_\odot \frac{R^2_0 + R^2_c}{R^2 + R^2_c},
\end{equation}
where $\rho_\odot$ is the local DM density at the Sun's galactocentric radius $R_0$, and $R_c$ is the core radius. The standard model values used are $\rho_\odot = 0.0079$ M$_\odot$ pc$^{-3}$, $R_0 = 8.5$ kpc, and $R_c = 5$ kpc. The circular rotation velocity is calculated using
\begin{equation}
V_\textrm{iso}(R) = \left[ 4\pi G \rho_0 R_c^2 \left( 1-\frac{R_c}{R} \tan^{-1}(R/R_c)\right) \right]^{1/2},
\end{equation}
where $\rho_0$ is the central DM density and is given by
\begin{equation}
\rho_0 = \frac{\rho_\odot}{1+\left(\frac{R_\odot}{R_c} \right)^2}.
\end{equation}
We keep $\rho_\odot$ and $R_c$ as free parameters for fitting.

\subsubsection{NFW model}

The NFW model allows more flexibility in the slope of the rotation curve than the semi-isothermal model does, since it is not restricted to producing flat rotation curves at large radii. The density profile is given by \citep{nfw1996}

\begin{equation} \label{eq:nfw}
\rho(R) = \frac{\rho_0}{\frac{R}{R_C} \left( 1 + \frac{R}{R_c} \right)^2},
\end{equation}
where
\begin{equation}
\rho_0 = \rho_\odot \frac{R_\odot}{R_c} \left( 1 + \frac{R_\odot}{R_c} \right)^2.
\end{equation}
The circular rotational velocity is given by \citep{nfw1996}
\begin{equation}
V_{\textrm{NFW}}(R) = \left[ 4\pi \rho_0 R_c^3 \left(\log(1+X) - \frac{X}{1+X}\right)\right],
\end{equation}
where $X = R/R_c$ \citep{sofue2012}. The fitting parameters in this model are $\rho_\odot$ and $R_c$.

\subsubsection{Power-law model}\label{sec:plm}

The power-law model is much more complicated in form than the semi-isothermal and NFW models, in part due to its ability to break spherical symmetry. Both the semi-isothermal and NFW models are spherically symmetric, but the power-law model can produce oblate and prolate halos. This does not actually make much of a difference in terms of the circular rotational velocity, as the symmetry axis of the halo is assumed to lie along the plane of the disk ($z=0$), thus any deviations away from spherical symmetry cancel out in the $+z$ and $-z$ directions. The density for this model is given by \citep{evans1993,evans1994}
\begin{equation}\label{eq:PLDH}
\rho(R, z) = \frac{v_0^2R_c^{\beta}}{4\pi G q^2} \frac{R^2_c(1+2q^2) + R^2(1-\beta q^2) + z^2 (2- (1+\beta) q^{-2})}{(R^2_c + R^2 + z^2q^{-2})^{(\beta+4)/2}},
\end{equation}
where $v_0$ is a normalisation velocity, $R_c$ is a scaling radius, $q$ is the ratio of the equipotentials, and determines whether the halo is a prolate ($q>1$) or oblate ($q<1$) spheroid, $\beta$ determines the asymptotic slope of the rotation curve ($\beta < 0$ gives a rising curve while $\beta > 0$ gives a falling curve). This is expressed in cylindrical coordinates, rather than spherical as the other two dark halo models are. The circular rotation velocity in the equatorial plane is computed using \citep{evans1993}
\begin{equation}
V_{\textrm{PL}}(R) = \left[ \frac{v_0^2 R^\beta_cR^2}{(R_c^2 + R^2)^{(\beta + 2)/2}} \right]^{1/2},
\end{equation}
where $v_0$, $\beta$, and $R_c$ are used as fitting parameters. The parameter $q$ cannot be physically constrained using rotation curves, for the reason explained above. Although $q$ does not affect the rotation curve fit, it will affect the optical depth along the line of sight, and therefore the expected number of events. 

\begin{figure}
\centering
\label{fig:df_plot}
\includegraphics[width=0.5\textwidth]{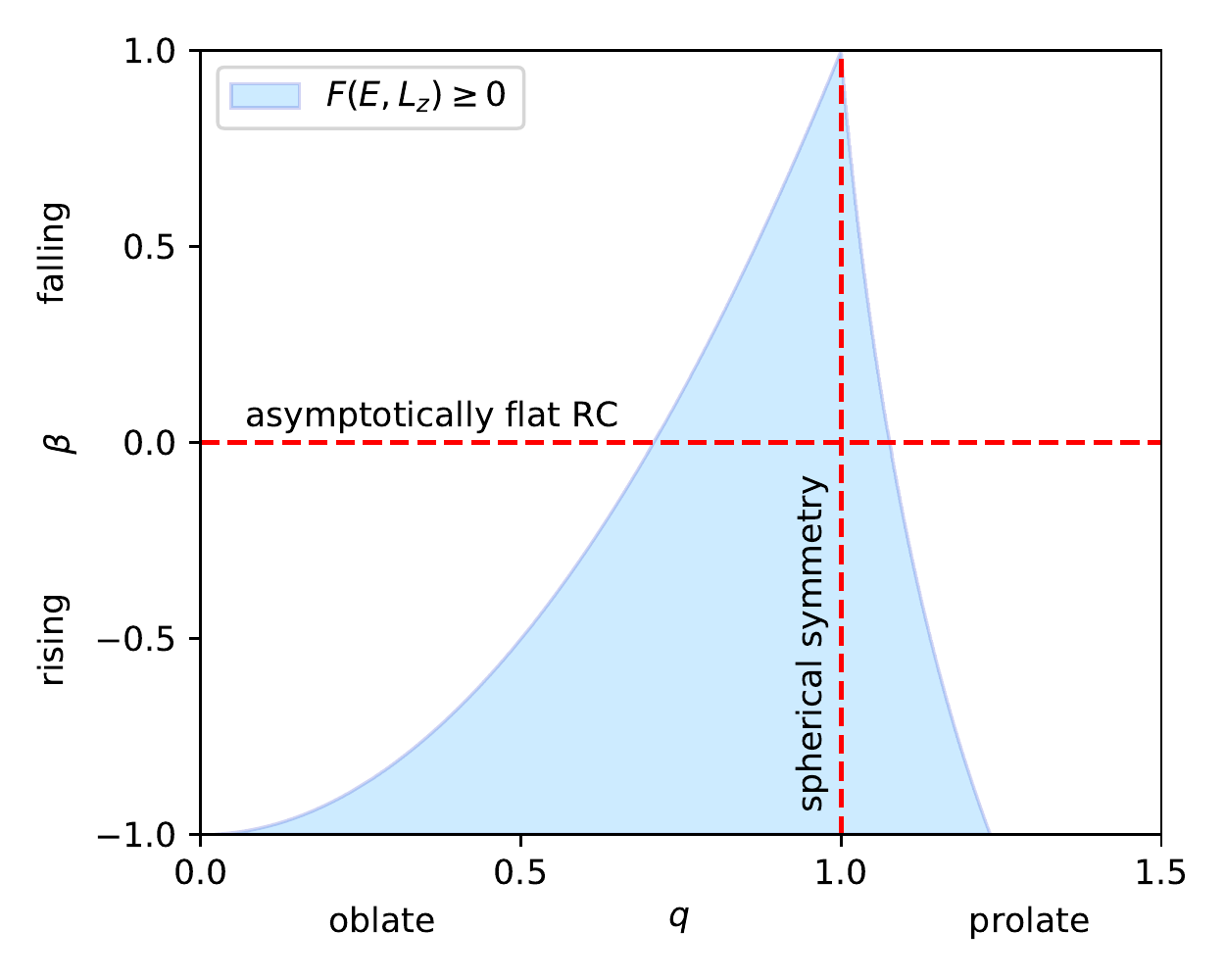}
\caption{The dependence of the phase-space density $F(E, L_z)$ on the parameters $\beta$ and $q$. The shaded regions represent the parameter space where $F(E, L_z)$ is positive definite, and hence is the allowed region. When the power-law model produces a falling rotation curve, the allowed region of halo shape decreases. The MWRC is best described by a model that produces a falling rotation curve at large radii. As a consequence of this, we will not be able to fully encapsulate the uncertainty in the shape of the dark halo using the power-law model.}
\end{figure}

Although the power-law model is much more flexible than both the isothermal and NFW profiles, it can still be rather limited in the shapes of haloes it can produce. The phase-space density, $F(E, L_z)$, which is a function of orbital energy $E$ and angular momentum $L_z$, must be positive definite for the model to be self-consistent. The condition for $F(E, L_z)$ to be positive definite is only dependent on the parameters $\beta$ and $q$ \citep{evans1993,evans1994}. In particular, when the rotation curve is declining (i.e. $\beta > 0$), the power-law model begins to demand a less oblate/prolate shape in order for the phase-space density to be positive definite, as seen in Figure \ref{fig:df_plot}. In the case where $\beta=1$, the only positive definite case is when $q=1$, that is, the model is spherically symmetric. 

When fitting the power-law model, we place a hard-wall prior such that $F(E, L_z) \geq 0$ to ensure that we are fitting a self-consistent dark halo model to the MWRC.
As a consequence of this, we can obtain model dependent constraints on the shape of the MW dark halo when fitting the MWRC. We stress that the constraints we obtain on $q$ are strictly model dependent, and do not necessarily reflect the true shape of the MW halo, but only reflect the possible shapes allowed by the power-law model.

\section{Milky Way Data and uncertainties}\label{sec:mwud}
In this section we outline some of the information and constraints on the Milky Way, including the MWRC dataset to be used in the fitting procedure, useful priors in constraining the mass models, and the possible shapes of the dark matter halo. One of the challenges of fitting the rotation curve is the relatively large degeneracy that occurs between each of the components that combine to form it. However, it will be shown that this degeneracy can be broken by including a prior on the baryonic mass in the disk model.

\subsection{Milky Way Halo Shape}\label{sec:mwq}

As discussed in section \ref{sec:plm}, the parameter $q$ cannot be efficiently constrained using rotation curve data alone. Accurate 3 dimensional motions of stars should provide a tight constraint on the shape of the dark halo. This will soon be a reality with forthcoming data from the \emph{Gaia} mission \citep{gaia2016}. For now it seems that there is very little agreement on the shape of the dark halo. Some argue that the shape is oblate \citep{olling2000, koposov2010, bowden2015}, prolate \citep{helmi2004, banerjee2011, bowden2016}, spherical \citep{fellhauer2006, smith2009}, and triaxial \citep{law2010,deg2013}.

Due to the requirements for the power-law model phase-space density to remain positive definite, we cannot test the full range of halo shapes. We explore the possible halo shapes that the power-law model allows. However, we note that our analysis will not be able to fully encapsulate the uncertainty in the halo shape, and will therefore underestimate the uncertainty in the microlensing event rate calculations.
The semi-isothermal and NFW models remain spherically symmetric, but the intrinsic uncertainty in the shape of the dark halo should be kept in mind.

\subsection{Local Stellar Surface Density Constraints}\label{sec:sfdc}

We consider the effect of adding a local stellar surface density constraint, $\Sigma_\odot$, that helps to break the degeneracy between the dark halo and disk models. The local stellar surface density is simply the amount of baryonic material (stars and gas) within a column with an area of one pc$^2$ at the galactocentric radius of the Sun. Many different estimates of the local stellar surface density exist and they are typically obtained by either stellar dynamics or by star counts (but also includes contributions from the interstellar medium). There is a general agreement between these two methods that $40 \lesssim \Sigma_\odot/M_\odot$pc$^{-2} \lesssim 60$. Constraints using stellar dynamics include $\Sigma_\odot = 48\pm 8M_\odot$pc$^{-2}$ \citep{kuijken1989}, $\Sigma_\odot = 51\pm 4M_\odot$pc$^{-2}$ \citep{bovy2013}, and $\Sigma_\odot = 44.4\pm 4.1 M_\odot$pc$^{-2}$ \citep{bienayme2014}. From the star counts side we have $\Sigma_\odot = 49.1 M_\odot$pc$^{-2}$ \citep{flynn2006}, $\Sigma_\odot = 54.2 \pm 4.9 M_\odot$pc$^{-2}$ \citep{read2014}, and $\Sigma_\odot = 47.1 \pm 3.4 M_\odot$pc$^{-2}$ \citep{mckee2015}. The dispersion in these results is much greater than the error bars quoted on each individual result.

Given the level of dispersion of these findings we opt to study two cases for a $\Sigma_\odot$ prior. 
We take the value obtained by \cite{bienayme2014} of $\Sigma_\odot = 44.4\pm 4.1 M_\odot$pc$^{-2}$ as a lower bound estimate, and take the value of $\Sigma_\odot = 54.2 \pm 4.9 M_\odot$pc$^{-2}$ by \cite{read2014} as an upper limit estimate. This way we can account for the dispersion in the in the local stellar surface constraints that exists in the literature.

\subsection{Dark Matter Substructure}

The dark matter halos that we consider in this paper produce smooth distributions of DM, and do not include any substructure. Further, we do not consider any possible LMC/SMC halos, which would add more DM mass between the observer and the sources we observe. In this subsection we wish to highlight these systematic uncertainties and some of their potential impact on MACHO constraints.

\subsubsection{Dark Matter Caustics}

It has been proposed that cold, collision-less DM can form caustics as it falls from all directions into a smooth gravitational potential \citep{sikivie1995, sikivie2003, sikivie2011, natarajan2007, Natsik2007, duffy2008}. The dark matter falls in towards the galactic centre, making its closest approach, and then converts its kinetic energy back into gravitational potential energy, and then repeats the process. 
Caustic rings form closer in towards the center of the galaxy at the turn-around points when the DM begins to turn its kinetic energy back into potential energy.

It is expected that this substructure will be destroyed during large galaxy mergers. However they may begin to form again if more DM begins to in-fall after a merger event, taking on the order of $10^8$ years to form \citep{dumas2015}. Some observational evidence for these DM caustics exists in other spiral galaxies \citep{kinney2000}, and in our own galaxy \citep{sikivie2003, dumas2015}.

If these DM substructure features, or any other form of DM structure, do indeed exist, not accounting for them will be a source of systematic uncertainty. Some caustic rings have been accounted for whilst fitting the MWRC \citep{deBoer2011,huang2016}, but this is done so by assuming a thin DM distribution for the rings, without any height scaling. Without an accurate representation of the three-dimensional profile this substructure follows, there is no way for us to accurately account for its effect on the optical depth of MACHOs along an arbitrary line-of-sight. The mass of this substructure should be small compared to the total mass of the dark halo, as was found in \cite{huang2016}. For this reason, we do not consider the DM caustics, or any other possible DM substructure, when deriving our MACHO constraints.

\subsubsection{LMC dark halo}

Measurements of the rotation curve of the LMC suggest that it contains a substantial fraction of DM \citep{vandermarel2002}. If this is indeed the case, not accounting for an LMC halo will mean that our constraints are under-estimated, as there will be more DM along a given line-of-sight to the LMC than expected based on the MW halo predictions. However in our analysis we wish to highlight the uncertainty on MACHO constraints due to the intrinsic uncertainties with the MW dark halo. Any additional uncertainty due to an LMC halo should be small compared to the that of the MW given the large difference in dark halo mass.

\subsection{Milky Way Rotation Curve datasets and fitting techniques}\label{sec:mwrctec}

\begin{table*}
    \centering
    \caption{The constraints for the isothermal and NFW halo models. Although the $\chi ^2$ per degree of freedom (d.o.f.) for the isothermal model is not significantly larger than 1, the disk model in the fit is compensating for the lack of flexibility from the dark halo model. Had we had placed a second prior on the disk shape to produce a realistic scale length, the resulting fit would be worse than reported here.}
    \label{tab:ISONFWmodel_params}
    \begin{tabular}{cccccccc }
        \hline
		Model & $M_b$ ($\times 10^{10}$\,M$_\odot$) & $a_b$ (pc) & $M_d$ ($\times 10^{10}$\,M$_\odot$) & $a_d$ (kpc) & $\rho_\odot$ ($\times 10^{-4}$\,M$_\odot$pc$^{-3}$) & $R_c$ (kpc) & $\chi ^2/$d.o.f. \\ 
		\hline
         ISO High $\Sigma_\odot$ prior & $0.907^{+0.088}_{-0.091}$ & $75.0^{+7.5}_{-7.5}$ & $4.59^{+0.47}_{-0.42}$ & $4.43^{+0.85}_{-0.83}$ & $98.3^{+4.8}_{-4.6}$ & $0.02^{+0.40}_{-0.00}$ & 1.13 \\ 
		ISO Low $\Sigma_\odot$ prior & $0.905^{+0.087}_{-0.093}$ & $74.6^{+7.9}_{-7.3}$ & $3.83^{+0.39}_{-0.35}$ & $3.95^{+0.81}_{-0.85}$ & $102.9^{+4.8}_{-4.5}$ & $0.02^{+0.51}_{-0.00}$ & 1.18 \\ 
		NFW High $\Sigma_\odot$ prior & $0.893^{+0.093}_{-0.087}$ & $74.7^{+7.7}_{-7.3}$ & $4.82^{+0.40}_{-0.48}$ & $2.96^{+0.25}_{-0.19}$ & $105.8^{+7.5}_{-7.2}$ & $13.4^{+3.0}_{-2.7}$ & 1.02 \\ 
		NFW Low $\Sigma_\odot$ prior & $0.894^{+0.091}_{-0.089}$ & $75.0^{+7.7}_{-7.5}$ & $4.09^{+0.41}_{-0.44}$ & $2.81^{+0.30}_{-0.19}$ & $114.2^{+6.9}_{-7.0}$ & $11.7^{+2.5}_{-2.3}$ & 0.976 \\ 
		\hline
    \end{tabular}
\end{table*}

For fitting, we decide to use the recent compilation by \cite{huang2016}, who uses data from the LSS-GAC \citep{liu2014,yuan2015}, SDSS/SEGUE \citep{segue2009}, and SDSS-III/APOGEE \citep{eisenstein2011,majewski2015} surveys to compile the MWRC. Other rotation curve compilations out to large galactocentric radius also exist in the literature, with \cite{xue2008}, \cite{bhat2014}, and \cite{sofue2013} being notable examples. The analysis by \cite{huang2016} contains newer releases of SDSS data and also a factor of $\sim$10$\times$ more stars than that of \cite{xue2008}, resulting in tighter errors in the resulting rotation curve.

A large difference between the analysis by \cite{bhat2014} and that of \cite{huang2016} is their treatment of the largely unknown velocity anisotropy parameter $\beta$, which relates the radial and transverse velocity dispersions. \cite{bhat2014} opt for using $\beta$ from simulations of the Milky Way, or choose arbitrary values \citep[e.g. see Fig 5 of][]{bhat2014}, whereas \cite{huang2016} use observationally determined values of $\beta$ and also propagate their associated uncertainties. See sec 4.2.2 of \cite{huang2016} for a comparison of their work with that of \cite{bhat2014}.

Since we are mostly interested in the MWRC at large galacto-centric radii, the bulge fit is not terribly important. Furthermore, the data by \cite{huang2016} does not probe the inner regions of the MWRC, and hence cannot be used to fit the bulge anyway. For these reasons we decide to fix the bulge model and propagate the uncertainty on the fixed values, in a similar fashion to \cite{mcmillan2011} and \cite{huang2016}. We set a prior of $a_b = 0.075 \pm 10\%$ kpc and $M_b = 8.9\times 10^9 \pm 10\% $ M$_\odot$. These values are consistent with previous estimates on the mass of the galactic bulge \citep{mcmillan2011, sofue2013, sofue2017}.

The galactocentric distance of the Sun is set to $R_0 = 8.34$ kpc, in line with recent observations by \cite{reid2014}. It should be noted that there is a considerable amount dispersion in the estimates for this distance. Some estimates place this distance as low as $\sim 6.7$ kpc \citep{branham2014}, while most of the literature points to  values with $R_0 \sim 8$ kpc \citep{bobylev2014, branham2015, branham2017, bajkova2016, vall2017, camarillo2018}. Should the true value of $R_0$ be different to the value used in this paper, we would expect a shift in our constraints on MACHOs. However, the uncertainty in $R_0 = 8.34 \pm 0.16$ kpc used to derive the rotation curve in \cite{huang2016} is propagated in their analysis, and is encapsulated in their uncertainties on the rotation curve. Therefore, when we propagate the uncertainties on our fits to the \cite{huang2016} data, some of the uncertainty in $R_0$ will be accounted for when we derive the constraints on MACHOs.

For the fitting procedure we employ a Markov Chain Monte Carlo (MCMC) sampling technique to explore the likelihood distribution of the data, and use the Python package \texttt{emcee} to employ it \citep{emcee2013}. The likelihood function is given by
\begin{equation}
L = \prod_{i=1}^N \frac{1}{\sqrt{2\pi} \sigma_i} \exp\left( -\frac{\left[ V^\textrm{obs}_{R_i} - V^\textrm{model}(R_i, \theta) \right]^2}{2 \sigma_i ^2} \right),
\end{equation}
where $N$ is the number of data points used in the fit, $\sigma_i$ is the uncertainty of the rotational velocity $V^\textrm{obs}_{R_i}$, and $\theta$ are the fitting parameters. When a Gaussian prior is added to the fit the likelihood function changes to
\begin{multline}
L = \frac{1}{\sqrt{2\pi} \sigma_P}\exp\left( -\frac{\left[ P^\textrm{obs} - P^\textrm{mod}(\theta) \right]^2}{2\sigma_{P}^2}\right) \\ \times \prod_{i=1}^N \frac{1}{\sqrt{2\pi} \sigma_i} \exp\left( -\frac{\left[ V^\textrm{obs}_{R_i} - V^\textrm{model}(R_i, \theta) \right]^2}{2 \sigma_i ^2} \right),
\end{multline}
where $P^\textrm{obs}$ is the prior, $P^\textrm{mod}$ is either the model for the prior or simply a free parameter, and $\sigma_{P}$ is the uncertainty on the constraint.

\subsection{Fitting Results}\label{sec:fitres}

\begin{table*}
     \centering
    \caption{The constraints for the power law dark halo models.}
    \label{tab:PLmodel_params}
    \begin{tabular}{cccccccccc}
        \hline
		Model & $M_b$ ($\times 10^{10}$\,M$_\odot$) & $a_b$ (pc) & $M_d$ ($\times 10^{10}$\,M$_\odot$) & $a_d$ (kpc) & $v_a$ (kms$^{-1}$) & $\beta$ & $R_c$ (kpc) & $q$ & $\chi ^2/$d.o.f. \\ 
		\hline
		High $\Sigma_\odot$ prior & $0.889^{+0.090}_{-0.090}$ & $74.8^{+7.7}_{-7.4}$ & $6.02^{+0.31}_{-0.35}$ & $2.49^{+0.10}_{-0.10}$ & $294^{+19}_{-24}$ & $0.73^{+0.15}_{-0.19}$ & $16.3^{+1.8}_{-2.2}$ & $0.995^{+0.018}_{-0.049}$ & 0.784  \\
		Low $\Sigma_\odot$ prior& $0.890^{+0.087}_{-0.092}$ & $75.1^{+7.4}_{-7.6}$ & $5.48^{+0.30}_{-0.32}$ & $2.325^{+0.092}_{-0.095}$ & $297^{+20}_{-24}$ & $0.71^{+0.16}_{-0.18}$ & $15.2^{+1.8}_{-2.1}$ & $0.995^{+0.018}_{-0.050}$ & 0.788
 \\ 
		\hline
    \end{tabular}
\end{table*}

\begin{figure*}
\label{fig:rot_plot}
\includegraphics[width=0.8\textwidth]{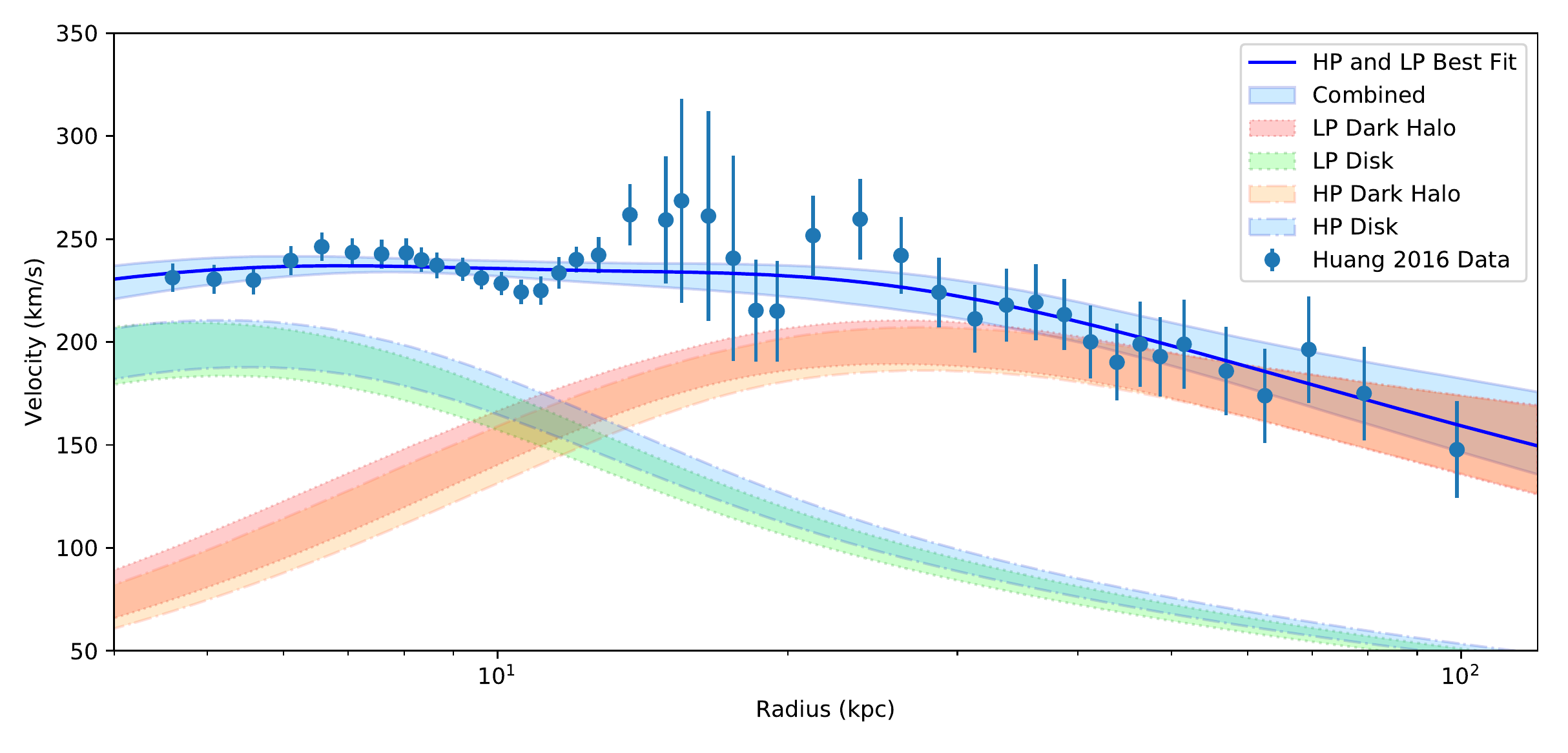}
\centering
\caption{The best fit plots and 2-$\sigma$ uncertainties (shaded) for the power-law model with both the low $\Sigma_\odot$ (LP) and high $\Sigma_\odot$ (HP) priors described in section \ref{sec:sfdc}. Our fixed model of the bulge is included in the combined and best fit constraints. The ``Combined'' shaded region is constructed using equation \ref{eq:vrot}. The difference between the combined rotation curve for the low and high prior cases is not discernible at the resolution of this plot, so we only show a single combined shaded region rather than one for each prior. The effect of the two priors is noticeable between the disk and dark halo contributions, where the high prior case makes a slightly more massive disk and therefore a slightly lighter dark halo in the inner regions of the rotation curve.}
\end{figure*}

\begin{table}
    \centering
    \caption{Model comparison tests for fits with the low and high $\Sigma _\odot$ prior, top to bottom. In either case, the isothermal dark halo model is strongly disfavoured compared to the power law dark halo model.}
    \label{tab:model_comp}
    \begin{tabular}{cccc}
        \hline
	Prior & Model & $\Delta$AIC & $\Delta$BIC \\ 
	\hline
	& ISO  &   10.4    &   8.7    \\ 
	Low $\Sigma_\odot$ & NFW  &    2.6    &    1.0    \\ 
	& PL   &    0.0    &    0.0   \\ 
	\hline
    & ISO  &   8.4    &   6.8     \\ 
	High $\Sigma_\odot$& NFW  &    4.3   &    2.7     \\ 
	& PL  &    0.0    &    0.0      \\ 
    \hline \hline
    \end{tabular}
\end{table}

The results for our fitting procedure are listed in Tables \ref{tab:ISONFWmodel_params} and \ref{tab:PLmodel_params} and the contour plots are presented in Figures \ref{fig:ISO}, \ref{fig:NFW}, and \ref{fig:PL}, for the semi-isothermal, NFW, and power-law models, respectively. The semi-isothermal model fits have some awkwardly shaped contours, and an unreasonably large fit for the disk scaling parameter $a_d$. The high value $a_d \sim 4$ kpc is quite inconsistent with photometrically derived values \citep{juric2008, licquia2016, bland2016, chen2017}. It appears that the disk component of the model is attempting to compensate for the lack of flexibility in the semi-isothermal halo model. This is not particularly surprising since the semi-isothermal model asymptotes to flat rotation curve at large radii, which is not evident in the observed MWRC data. 

We now briefly discuss previous constraints on the Milky Way thin disk.
Using 48 million stars from SDSS, \cite{juric2008} produced stellar density maps of the Milky Way. They then fit the stellar density maps to obtain photometrically derived constraints on the length scale of the Milky Way thin and thick disks \citep{juric2008}. They find a thin disk length scale of $\sim 2.5 - 3 $ kpc, which is not tightly constrained due to SDSS observing high galactic latitudes. Using the newer SDSS data, in addition to data from XSTPS-GAC, \cite{chen2017} constrain the thin disk length scale to $\sim 2.35 $ kpc. By performing a Bayesian meta-analysis of 29 different photometrically derived values of the thin disk scale length, \cite{licquia2016} constrain the value to $2.64 \pm 0.13$ kpc. We have focused on the thin disk length scale in this brief discussion since it is the more dynamically relevant disk component, containing roughly 5$\times$ more mass than the thick disk \citep{bland2016,huang2016}.

The NFW model performs better than the semi-isothermal model since it is not confined to producing flat rotation curves. However, we needed to add in a hard wall prior of $a_d < 4$ kpc to prevent the disk model from fitting the dark halo dominated regions of the MWRC, and vice-versa. 

The power law model is by far the most flexible of the dark halo models tested. It is able to produce a sensible fit even without any priors, but we still add these in to better reflect the current uncertainty in the local stellar surface density. The somewhat triangular shape of the contours of the $q$ parameter are a result of the requirement that the phase-space density be positive-definite in each step of the MCMC. Since MWRC is falling at larger $R$, the $\beta$ parameter gets constrained such that $\beta > 0$. The triangular shape comes from the top shaded region in Figure \ref{fig:df_plot}.

Figure \ref{fig:rot_plot} shows the $2\sigma$ uncertainty regions for the fits with the low and high $\Sigma_\odot$ priors. The uncertainty in the disk and dark halo models is much larger than the combined rotation curve, owing to the large degeneracy that exists between these two models. Only a slight difference in the resulting fit arises from the choice of prior used.

To compare the three dark halo models we have tested, we use Information Criteria (IC) tests. The Akaike Information Criteria \cite[AIC;][]{aic1974} and the Bayesian Information Criteria \cite[BIC;][]{bic1978} are the two we choose to use, and are given by
\begin{align}
AIC &= -2\ln L + 2k, \\
BIC &= -2\ln L + k\ln N,
\end{align}
where $k$ is the number parameters in the model, and $N$ is the number of data points. These criteria encapsulate Occam's razor, thus models with more parameters are penalised. The model with the lowest AIC or BIC is considered better at explaining a given dataset. A difference in AIC or BIC of 2 is considered positive evidence in favour of the model with the lower IC, while a difference of 6 is considered strong evidence \citep{kass1995}.

The results for each IC test are presented in Table \ref{tab:model_comp}. We can see that the semi-isothermal model is strongly disfavoured compared to both the NFW and power law models, particularly for the low $\Sigma_\odot$ prior. This would be due to the fact that the isothermal model is producing a flat rotation curve, and a large disk is better at producing a falling rotation curve at large radii.
While the NFW model does a better job than the isothermal model, it still does more poorly than the power law model.  The information criteria tests show that the extra parameters in the power law model are justified, as they lead to significant improvement in the quality of the fit.

The main result we wish to illustrate here is that the semi-isothermal model does not produce an adequate fit to the MWRC. Despite this, it has been the most commonly used model for determining microlensing rates in the Milky Way halo. The power law model provides the most robust fit to the data, and is therefore a suitable replacement.

\section{Microlensing Formalism}\label{sec:micro}
\subsection{Microlensing Amplification}
PBHs can be detected by their gravitational influence as they pass very close to the line of sight of a distant star \citep{pacz1986}. Just as in large scale gravitational lensing, a microlensed source will be distorted into multiple images as the lens passes by. If the lens and source are perfectly aligned, an Einstein ring of radius
\begin{equation}
r_E(M, x) = \sqrt{\frac{4GMLx(1-x)}{c^2}}
\end{equation}
is formed, where $M$ is the mass of the lens, $L$ is the distance between the observer and source star, and $x$ is the ratio of the observer-lens and observer-source distances. In reality it is very unlikely that the lens will fall perfectly on the line-of-sight to the source, so multiple images of the source, rather than a ring, will be produced. These multiple images are not resolvable in the case of microlensing. However, what will be detectable is the apparent brightening caused by the multiple line-of-sights now able to reach the observer. An apparent amplification of
\begin{equation}\label{eq:amp}
A = \frac{u^2 + 2}{u\sqrt{u^2 + 4}},
\end{equation}
will be observed, where $u=b/r_E$ is the impact parameter in units of the Einstein radius, and $b$ is the perpendicular separation between the lens to the observer-source line of sight.

\subsection{Optical Depth}

The optical depth $\tau$ is defined as the number of compact lenses within a tube of radius $u_Tr_E$ and length $L$, where $u_T = 1$ is the threshold parameter and corresponds to a minimum brightening of the source star by $A_T \sim 1.34$. The optical depth is a useful parameter in that it is independent of the MACHO mass $M$, and only depends on the density along the line of sight 
\begin{equation}
\tau = \int \frac{\rho(s)}{M} \pi u_T^2 r_E^2(s) ds,
\end{equation}
where $\rho(s)$ is the density along the line of sight.

\subsection{Microlensing Rate}
The microlensing rate $\frac{d\Gamma}{d\hat{t}}$, where $\hat{t}$ is the time it takes for the lens to cross the Einstein ring,\footnote{Note that MACHO uses $\hat{t}$ to refer to the time it takes for the lens to cross the Einstein \emph{ring}, where was EROS-2 has $t_E$ as the time it takes the lens to cross the Einstein \emph{radius}. Thus $\hat{t} = 2t_E$.} is a model dependent function that allows one to calculate the expected distribution of event durations. This can then be directly used to determine the expected number of microlensing events one should observe assuming some fraction $f$ of the halo is comprised of PBHs. 

\subsubsection{Semi-isothermal}
For the semi-isothermal dark halo model, \cite{alcock1996} obtain
\begin{equation}\label{eq:rate}
\frac{d\Gamma}{d\hat{t}} = \frac{512Lu_T\, G^2M\rho_0(R_\odot^2+R_c^2)}{\hat{t}^4v_c^2c^4} \int^{x_h}_0 \frac{x^2(1-x)^2}{A+Bx+x^2} e^{-Q(M,x)}dx,
\end{equation}
where $v_c$ is the galactic rotation velocity of the Sun, $x_h \approx 1$ is the extent of the halo, and
\begin{align}
Q(M,x) &= \frac{4r_E^2(M,x)\,u_T^2}{\hat{t}^2v_c^2}, \hspace{1cm}
A = \frac{R_c^2 + R_\odot^2}{L^2}, \\
B &= -2\frac{R_\odot}{L}\cos b\cos l,
\end{align}
where $b$ and $l$ are the Galactic latitude and longitude respectively.

\subsubsection{NFW}
For the NFW dark halo profile we derive
\begin{equation}
\begin{aligned}\label{eq:rate2}
&\frac{d\Gamma}{d\hat{t}} = \frac{512L^2u_T\, G^2M\rho_0R_c}{\hat{t}^4v_c^2c^4}\times \\ &\int^{x_h}_0 \frac{x^2(1-x)^2 e^{-Q(M,x)}}{\left[A'+Bx+x^2\right]^{1/2}\left( 1 + \frac{L}{R_c}\left[A'+Bx+x^2\right]^{1/2} \right)^2} dx,
\end{aligned}
\end{equation}
where
\begin{equation}
A' = \frac{R_\odot^2}{L^2}.
\end{equation}
The derivation of equation (\ref{eq:rate2}) can be performed starting with equation (A2) from \cite{alcock1996} with the NFW model expressed in terms of the density along the line of sight from the Solar position, rather the from the galactic centre as it is in equation (\ref{eq:nfw}).
\subsubsection{Power-law}

Finally, for the power-law model, the microlensing rate is given by
\begin{multline}\label{eq:PL}
\frac{d\Gamma}{d\hat{t}} = 8\frac{u_T}{\pi c^2} \left( \frac{L^6}{R_c^4\hat{t}^4} \right) (\beta + 2) |\beta|^{1+4/\beta}\frac{1-q^2}{q^2}\Big(a_1 G' J_1 + a_2 H' J_2\Big)\ \\ \hspace*{1cm}\ + 8\frac{u_T}{\pi c^2} \left( \frac{L^4}{R_c^2\hat{t}^4} \right) \frac{(\beta + 2) |\beta|^{1+4/\beta}}{q^2}  a_1 J_1 \\ \hspace*{-2.5cm}+ 8\frac{u_T}{\pi c^2} \left( \frac{L^4}{R_c^2\hat{t}^4} \right) |\beta|^{1+2/\beta} \Big[2-\frac{1+\beta}{q^2}\Big] a_3 J_3,
\end{multline}
where the parameters are defined in Appendix B of \cite{alcock1995}.

\subsection{Wide mass distributions}

When the mass distribution of PBH is no longer monochromatic,\footnote{In this paper we use monochromatic meaning of a single mass, rather than of a single colour. This is common parlance in the field \citep[e.g. see ][]{carr2010, carr2016, 7hints2017}.} the equation for the microlensing rate can be generalised to take in any arbitrary mass distribution. In the case of the semi-isothermal dark halo model (\ref{eq:rate}), this becomes
\begin{eqnarray}\label{eq:int}
\frac{d\Gamma}{d\hat{t}} &=& \frac{512Lu_T\, G^2\rho_0(R_\odot^2+R_c^2)}{\hat{t}^4v_c^2c^4} \times \\ \nonumber
&&\hspace{-1cm}\int_0^\infty dM\,\psi(M,\mu,\sigma)\,M \int^{x_h}_0 dx\,\frac{x^2(1-x)^2}{A+Bx+x^2} e^{-Q(M,x)}\,,
\end{eqnarray}
and analogously in (\ref{eq:rate2}) and (\ref{eq:PL}).
The distribution of PBH masses produced by inflationary models is well approximated by a normalized lognormal distribution \citep{green2016}
\begin{equation}\label{eq:LN}
\psi(M,\mu,\sigma) = \frac{f_{\rm PBH}\ e^{-\frac{1}{2}\sigma^2}}{\mu\,\sqrt{2\pi\sigma^2}}\, \exp\left(-\frac{\ln^2M/\mu}{2\sigma^2}\right)\,,
\end{equation}
where $\mu$ is the peak of the distribution, and $\sigma$ is its dispersion (i.e. width).
The first two reduced moments of the lognormal distribution (\ref{eq:LN}) are, see~\cite{juan2017},
\begin{equation}\label{eq:moments}
\langle M\rangle = \mu\,e^{\frac{3}{2}\sigma^2}\,,  \hspace{6mm}
\Delta M^2 \equiv \langle M^2\rangle - \langle M\rangle^2 = \langle M\rangle^2 \left(e^{\sigma^2}-1\right)\,.
\end{equation}

\subsection{Clustered PBH}
The effect of clustering, with $N_{\rm cl}$ PBH per cluster, is simply to change the reduced moments of the distribution,
\begin{equation}\label{eq:equiv}
\langle M\rangle_{\rm equiv} = N_{\rm cl} \, \langle M\rangle\,,  \hspace{5mm}
\Delta M^2_{\rm equiv} = N_{\rm cl}^2 \, \Delta M^2\,,
\end{equation}

which means that the mean and dispersion parameters of the equivalent lognormal distribution change from $(\mu,\,\sigma)$ to $(\tilde\mu,\,\tilde\sigma)$ with, see~\cite{juan2017},
\begin{equation}\label{eq:newms}
\tilde\mu = N_{\rm cl}\,\mu\ e^{\frac{3}{2}(\sigma^2-\tilde\sigma^2)}\,,  \hspace{6mm}\tilde\sigma^2 = \ln\left(1 + \frac{e^{\sigma^2}-1}{N_{\rm cl}}\right)\,.
\end{equation}
In this case, we constrain the peak in the distribution of PBH cluster masses, $\tilde\mu$. The microlensing constraints will then shift towards smaller values due to the integral in equation~(\ref{eq:int}). Models of PBH with a wide mass distribution $\psi(M)$ that were ruled out by the microlensing surveys are still allowed when PBH clustering is considered. Very modest values of $N_{\rm cl}\sim10-100$ are enough.

\subsection{Expected Number of Events}

The expected number of events is given by
\begin{equation}
N_\textrm{exp} = E \int_0^\infty \frac{d\Gamma}{d\hat{t}} \xi (\hat{t}) d\hat{t},
\end{equation}
where $E$ is the total exposure time in units of star years (that is, if we look at $10^6$ stars for one year, $E=10^6$ star years), and $\xi (\hat{t})$ is the efficiency function. The efficiency function encapsulates the ability of a given survey to identify microlensing events over a range of $\hat{t}$. If a survey only runs for a short period of time, it will not be able to efficiently detect long duration microlensing events.

The long duration events are primarily caused by more masses lenses. The reason for this is two-fold. Firstly, a higher mass lens has a larger Einstein radius, and therefore a larger lensing footprint. Secondly, in a virialised system, more massive objects tend to move more slowly than less massive objects, and will therefore take longer to cover the same distance. These effects combine to limit the sensitivity of massive lenses by microlensing surveys.

\subsection{Obtaining Microlensing Constraints}

From the expected number of events we can easily obtain the fractional constraint, $C(M)$, on PBH DM using
\begin{equation}
C(M) = \frac{N_\textrm{avg}}{N_\textrm{exp}(M)},
\end{equation}
where $N_{\textrm{avg}}$ is the upper limit on the average event rate of microlensing events, obtained from the Poisson distribution.
The Poisson distribution of microlensing events is given by
\begin{equation}
P(N_{\textrm{obs}}) = e^{-N_\textrm{avg}} \frac{N_\textrm{avg}}{N_{\textrm{obs}}!},
\end{equation}
where $N_\textrm{obs}$ is the number of observed microlensing events for a given survey. We want to find the probability such that $P(N_\textrm{obs}) = 0.95$ for some upper $N_\textrm{avg}$. That is, we want to find the value of $N_\textrm{avg}$ where we can be 95\% confident that the true average event rate is not larger than $N_\textrm{avg}$. For the EROS-2 survey, no events were observed, so we can be 95\% confident that true average event rate is not larger than 3 events. For a given value of $N_{\rm avg}$, the condition
\begin{equation}
C(M) \leq 1
\end{equation}
constrains PBH at mass $M$ from contributing 100\% of the DM halo.

\subsection{Wide mass distributions and clustered PBH Constraints}\label{sec:WMD}

The final microlensing constraints when considering a wide mass distribution can be obtained by simply integrating out the monochromatic distribution constraint, see (\ref{eq:int}), as
\begin{equation}\label{eq:wide}
\int_0^\infty dM\,\psi(M,\mu,\sigma)\,N_{\rm exp}(M) \leq N_{\rm avg}\,,
\end{equation}
which in the case of a lognormal distribution (\ref{eq:LN}) can then be written as constraints on $f_{\rm PBH}$ as a function of the $\mu$ parameter, for different values of $\sigma$
\begin{equation}\label{eq:wide2}
f_{\rm PBH}(\mu) \leq \left[\int_0^\infty dM\,\frac{\bar\psi(M,\mu,\sigma)}{C(M)}\right]^{-1}\,,
\end{equation}
where $\psi=f_{\rm PBH}\bar\psi$, see (\ref{eq:LN}). 

\begin{figure}
\centering
\label{fig:eros_ext_mass}
\includegraphics[width=0.5\textwidth]{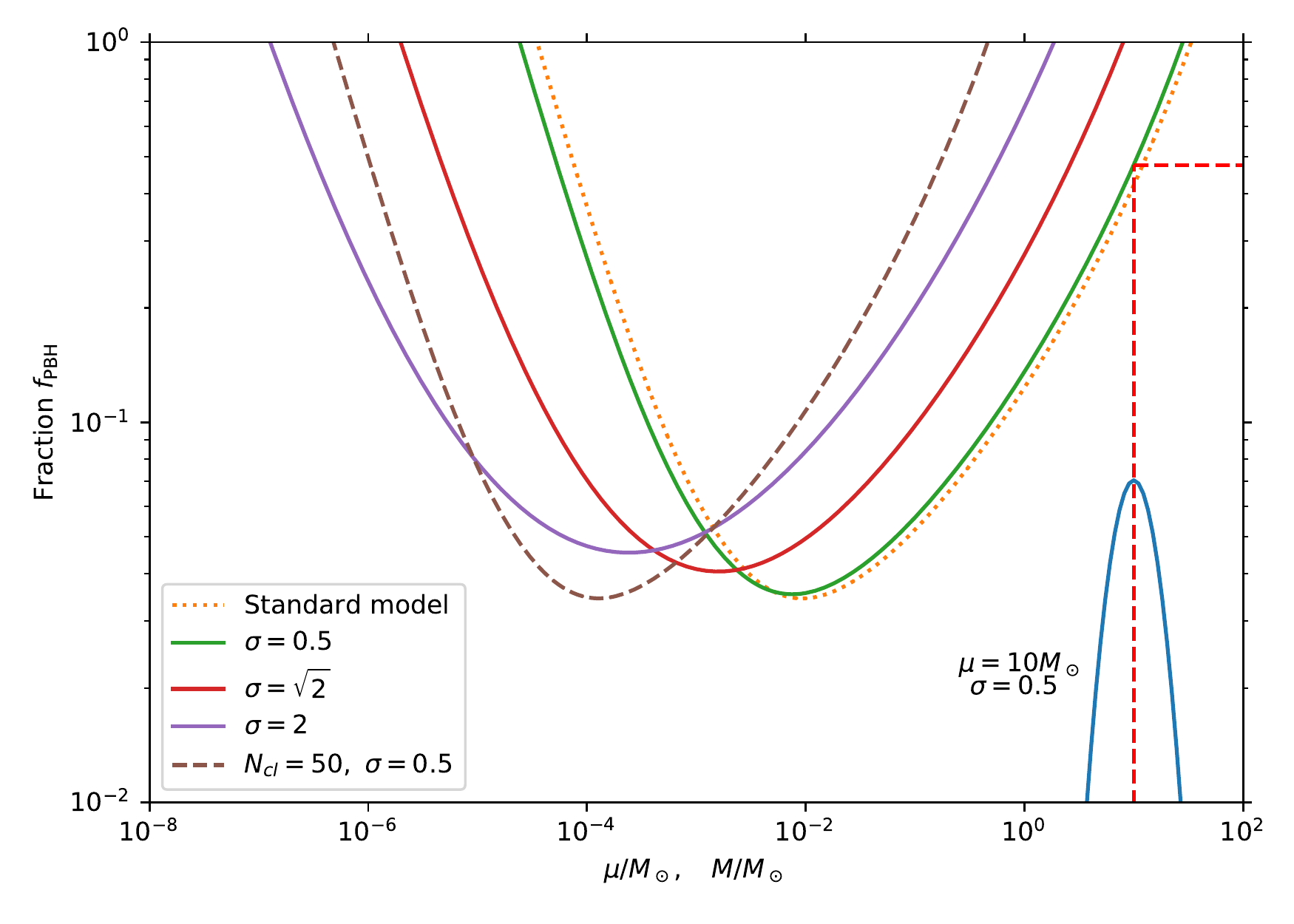}
\caption{The EROS-2 standard model constraints for an extended mass function and clustered PBHs. For $\sigma=0.5$, the constraints closely follow the monochromatic mass constraints (orange dotted line), however the constraints shift to lower masses for broader distributions. In the bottom right, an example of a lognormal distribution with parameters $[\mu,\, \sigma]=[10M_\odot,\, 0.5]$ is plotted. A vertical line drawn from the peak of the distribution intercepts the $\sigma = 0.5$ constraint (in green, right solid line) at $f_{\textrm{PBH}}=0.475$, meaning that this distribution is excluded from contributing more than 47.5\% of DM at 95\% c.l. A PBH distribution with $[\mu,\, \sigma]=[10M_\odot,\, 2]$, or even $[\mu,\, \sigma]=[10M_\odot,\, \sqrt{2}]$, could still make up the entirety of DM.}
\end{figure}

We show in Figure \ref{fig:eros_ext_mass} the effect of a wide mass distribution on the generic microlensing constraints from the EROS-2 survey. We also show the effect of clustering on the same constraints. In the case of the wide mass distribution, the constraints shift towards lower masses since for a wider mass distribution, there are progressively more massive lenses which can be more difficult to detect as they require a longer microlensing survey duration to detect. When the width of the distribution is small, as is mostly the case for $\sigma=0.5$, the lognormal distribution is well approximated by a delta function.

For the clustered PBH, the constraints essentially change from $C(M)$ into $C(N_{\rm cl} \mu e^{\frac{3}{2} \sigma^2 })$, since for large $N_{\rm cl}$ the new clustered distribution is well approximated by a delta function (i.e. $\tilde\sigma << 1$), provided that $\sigma$ is not too large. 

The way to interpret Figure \ref{fig:eros_ext_mass} is the following. For a given lognormal dispersion $\sigma$, compute the integral (\ref{eq:wide2}) for an array of peak masses $\mu$. This gives the upper bound on a set of lognormal populations with the chosen set of $\mu$, and a fixed $\sigma$. We show three different values of $\sigma$ plotted in Figure \ref{fig:eros_ext_mass}, for an array of peak masses $\mu$ on the $x$-axis.

These three curves give the 95\% confidence interval that a PBH population with a given $\mu$ and $\sigma$ is excluded from making up more than $f_\textrm{PBH}$. We demonstrate such a distribution in Figure \ref{fig:eros_ext_mass}, where the lognormal distribution with parameters $[\mu,\, \sigma]=[10M_\odot,\, 0.5]$ is plotted. The vertical line that crosses the peak of the distribution hits the $\sigma=0.5$ constraint (green line) at $f_{\textrm{PBH}}=0.475$, meaning that the distribution with $[\mu,\, \sigma]=[10M_\odot,\, 0.5]$ is excluded from making up any more than 47.5\% of DM at the 95\% confidence interval.

For a population not to be excluded, the peak of the distribution must not intercept the line for $f_\textrm{PBH}\leq 1$ in the way described in the previous paragraph. For example, if we had plotted a lognormal distribution with $[\mu,\, \sigma]=[10M_\odot,\, 2]$, or even $[\mu,\, \sigma]=[10M_\odot,\, \sqrt{2}]$, the peaks of both distributions would not intercept their respective exclusion regions for $f_\textrm{PBH}\leq 1$ (i.e. the purple and red lines, respectively). Note that the mean mass of the lognormal distribution is much larger than the peak mass (the mode). It would be equivalent to produce the plots as a function of the mean mass rather than the mode of the distribution, but the visual interpretation would be more cumbersome since the mean is strongly displaced from the peak for large $\sigma$.

\begin{figure}
\centering
\label{fig:eff}
\includegraphics[width=0.45\textwidth]{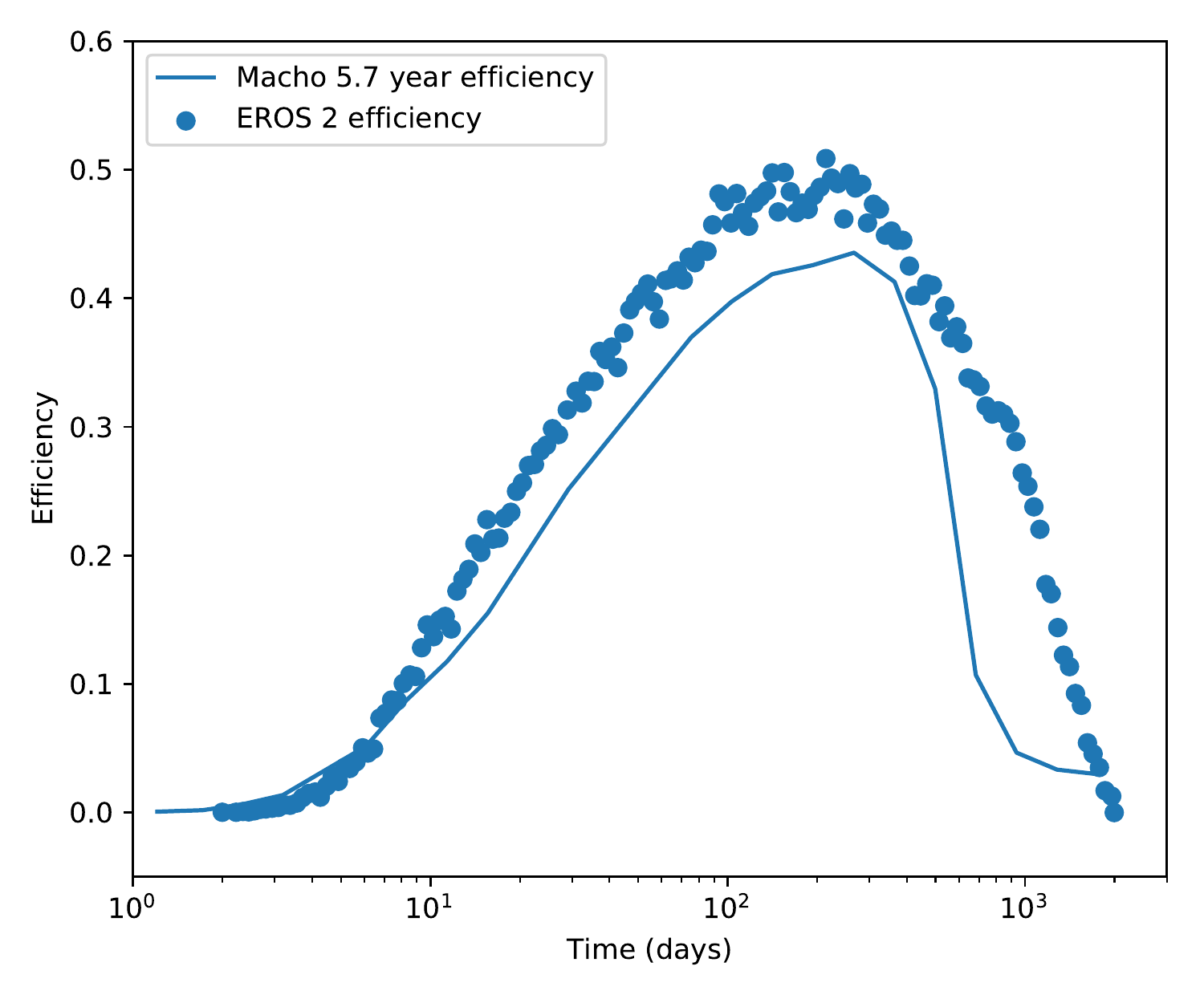}
\caption{The efficiency functions for the MACHO Collaboration 5.7 year (criteria A) and EROS-2 surveys, showing the efficiency with which each survey could detect lenses as a function of the Einstein ring crossing time \citep{alcock2000,eros2007}.}
\end{figure}

\section{Results and Discussion}\label{sec:res}

We consider two constraints on the MACHO fraction in the halo, the MACHO Collaboration 5.7 year \citep{alcock2000} and the EROS-2 LMC bright star sample results \citep{eros2007}. The efficiencies of both surveys are plotted in Figure \ref{fig:eff}. The efficiency functions attempt to reflect the ability of each survey to detect genuine microlensing events. There are uncertainties associated with these efficiency functions, some quantifiable and others not. For example, the MACHO Collaboration estimate that the uncertainty of their efficiency function could be on the order of $\sim 20$\% \citep{alcock2001}, and comes mostly from blending of the source stars. Unquantifiable uncertainties include lensing from exotic lenses, which would result in oddly shaped light-curves, as might be expected from spatially clustered sources.

In stark contrast to the lack of events observed by EROS-2, the MACHO Collaboration found a total of 17 events using their selection Criteria B, and 13 events using their criteria A \citep{alcock2000}. This is despite being quite similar in their efficiencies, and only about a factor of two different in their total exposure times.

This contrast leads to drastically different constraints on the optical depth of MACHOs towards the LMC. The MACHO Collaboration find that $\tau = 1.2^{+0.4}_{-0.3} \times 10^{-7}$, with an additional 20\% to 30\% systematic uncertainty, while the EROS-2 survey finds an upper limit of $\tau < 0.36 \times 10^{-7}$ at the 95\% confidence interval. Although we will not be updating their analysis, it is worth mentioning that the microlensing constraints towards the LMC obtained by the OGLE Collaboration are in agreement with the constraints obtained by the EROS Collaboration \citep{ogle2011}.\footnote{The main reason we do not update the OGLE constraints is that it is not possible to reproduce their microlensing constraints based on the information they provide in the paper detailing their analysis \citep{ogle2011}, as they do not provide all of the efficiency functions for their fields, or an average efficiency function as MACHO and EROS do.} The disagreement between the optical depth constraints between the MACHO and EROS Collaborations has received the attention from numerous authors \citep[e.g.\ see][]{nelson2009, besla2013}, but motivates further study.

Follow up observations of the microlensed stars from the MACHO Collaboration revealed that some of them had repeated variability in their light curves, which strongly disfavours a microlensing interpretation of the original variability. With newer information, \cite{bennett2005} updates the MACHO 5.7 year results to find that $\tau = (1.0 \pm 0.3) \times 10^{-7}$, which is still not in good agreement with the constraints from EROS-2. 


\begin{figure}
\centering
\label{fig:er_nevs}
\includegraphics[width=0.5\textwidth]{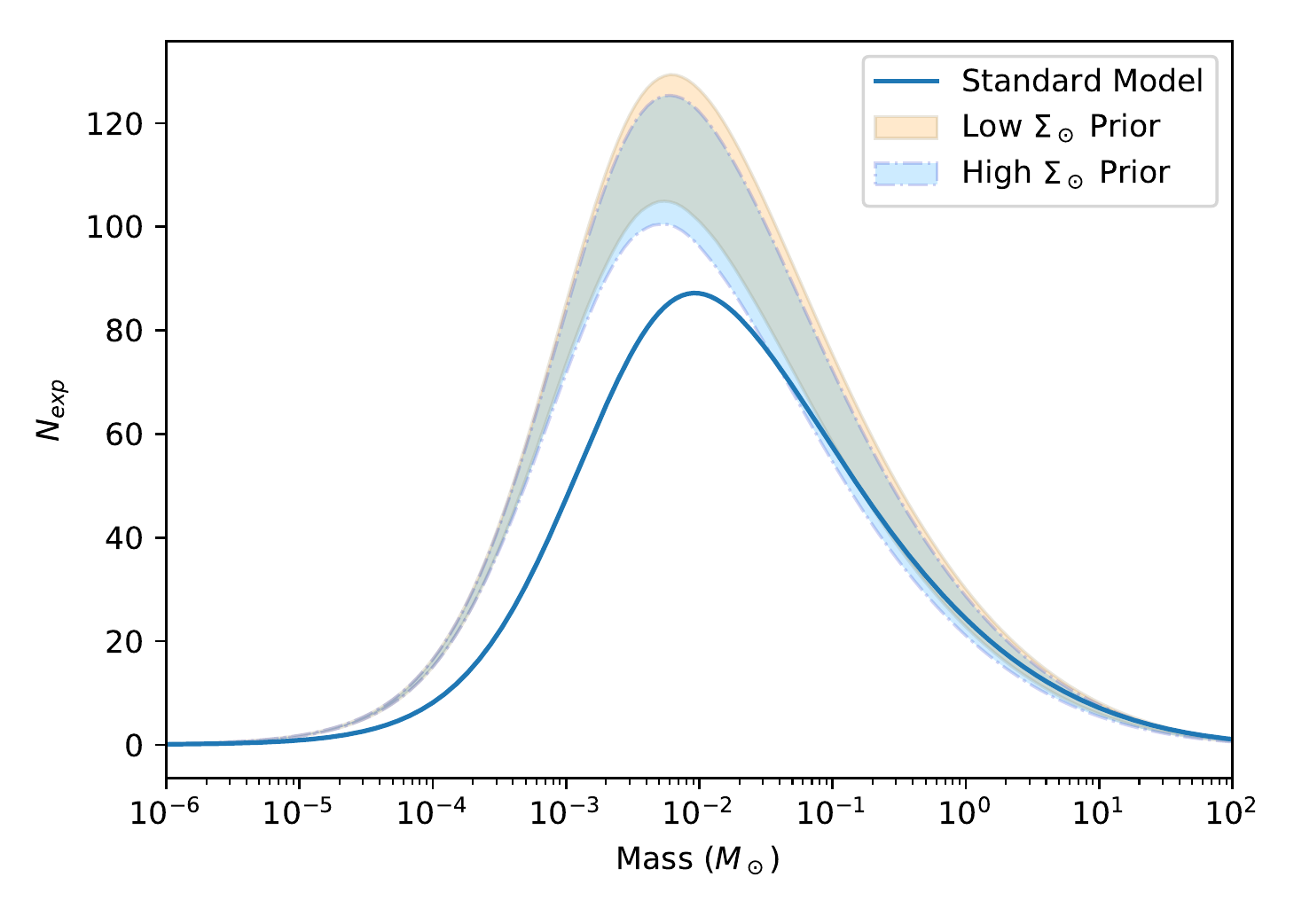}
\caption{The expected number of microlensing events the EROS-2 survey should have seen if all DM was in the form of MACHOs with some mass $M$. The shaded regions represent the 2-$\sigma$ uncertainty propagated from the MWRC fit in Figure \ref{fig:rot_plot}, and also includes the uncertainty in the shape of the MW dark halo. Here we have plotted the distribution in the expected number of events with both the low and high $\Sigma_\odot$ prior that was used in fitting the MWRC. As one might expect, the low $\Sigma_\odot$ prior on the disk fit results in a larger dark halo, which results in a higher expected number of events.}
\end{figure}

\begin{figure}
\centering
\label{fig:er_nevsm}
\includegraphics[width=0.5\textwidth]{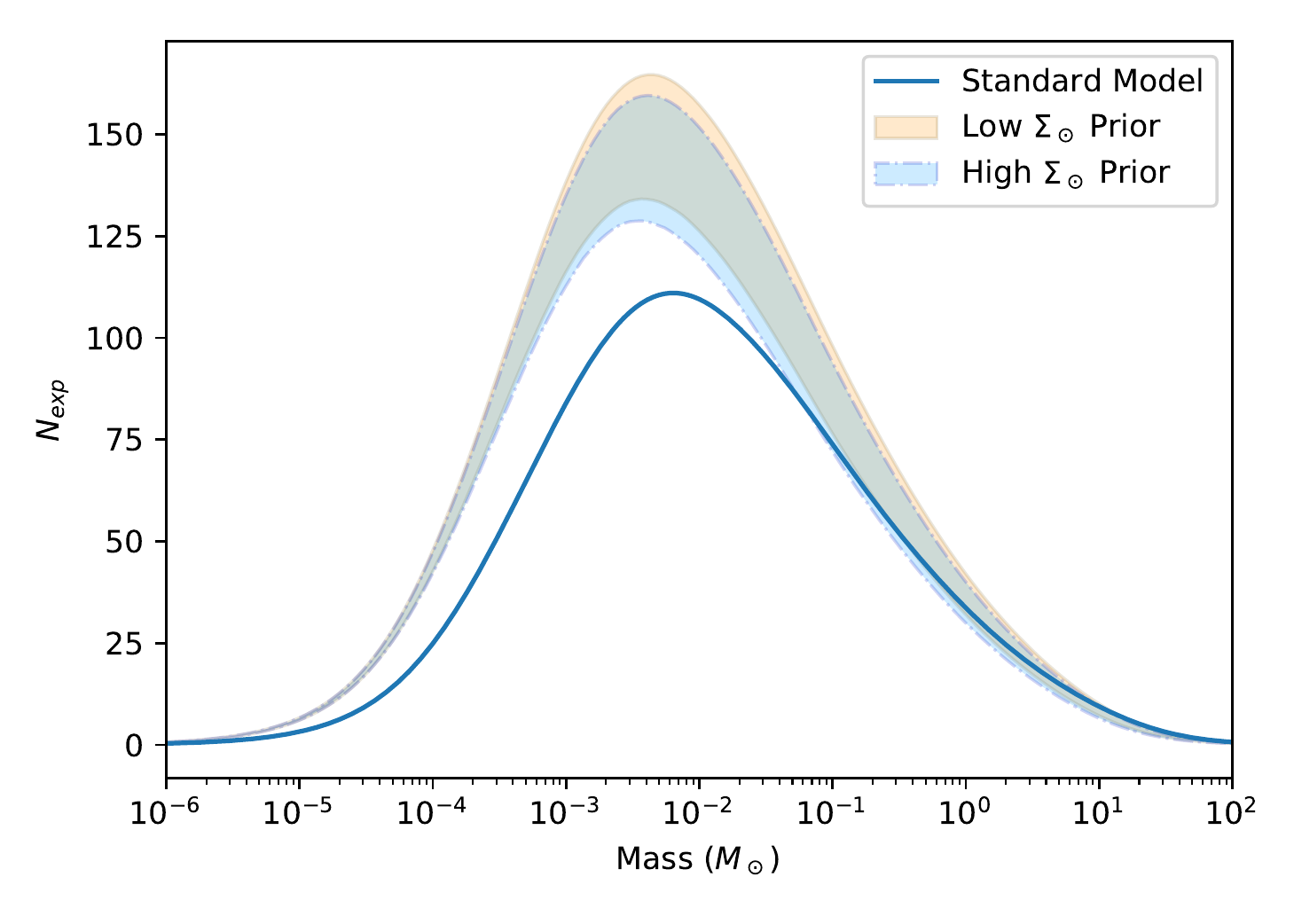}
\caption{The same as Figure \ref{fig:er_nevs}, but using the MACHO Collaboration 5.7 year efficiency.}
\end{figure}

With a new model for the MW dark halo selected (see section \ref{sec:fitres}), we can now move onto deriving constraints on the MACHO fraction of the dark halo using both the MACHO Collaboration 5.7 year and the EROS-2 results and efficiency functions. We will also propagate the uncertainties from the MWRC fits whilst rederiving the constraints for each survey.

To propagate the uncertainty, we sample steps from our MCMC chains and compute a microlensing model and event rate for a range of lens mass $M$ values, using the parameters $[v_{a_i}, \beta_i, R_{c_i}, q_i]$, where $i$ denotes the $i$-th step in the MCMC chain. The result of this is that for each lens mass $M$ we will have a range of models with parameters $[v_{a_i}, \beta_i, R_{c_i}, q_i]$ that reflects the range of models allowed by the MWRC fits from earlier. From this range of models, we can then compute the 2-$\sigma$ uncertainty values on the expected number of microlensing events, $N_\textrm{exp}$, for each lens mass $M$. That is, for each $M$, we obtain a range of $N_\textrm{exp}$ that is consistent with the MWRC data. The resulting uncertainty regions are plotted in Figure \ref{fig:er_nevs}. The uncertainty values are obtained using the cumulative distribution function of the range of expected number of events for each lens mass $M$.

\begin{figure}
\centering
\label{fig:tau_los}
\includegraphics[width=0.5\textwidth]{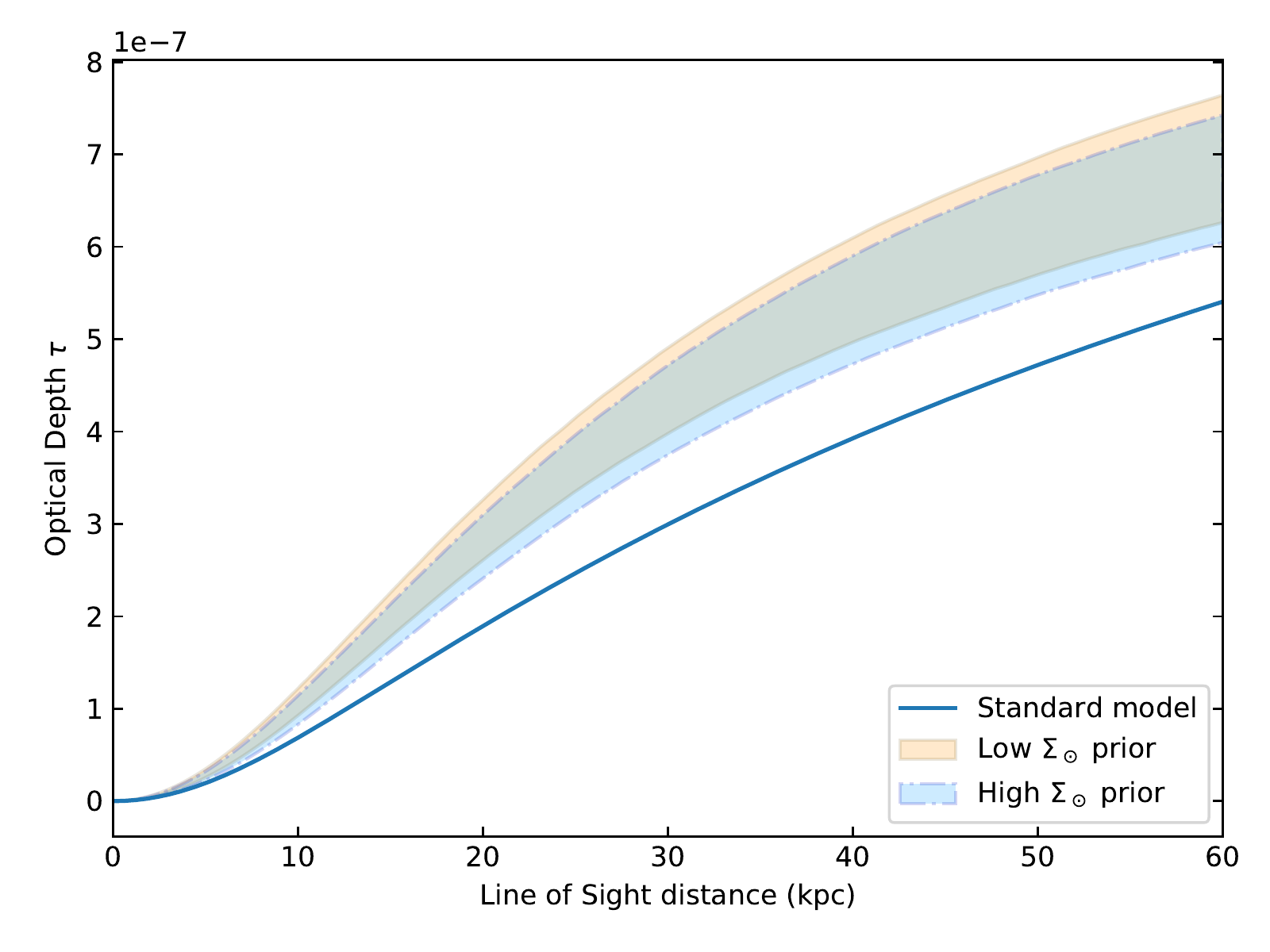}
\caption{The optical depth $\tau$ along the line of sight towards the LMC for the low and high $\Sigma_\odot$ prior, as well for the standard model.}
\end{figure}

\begin{figure}
\centering
\label{fig:er_up}
\includegraphics[width=0.5\textwidth]{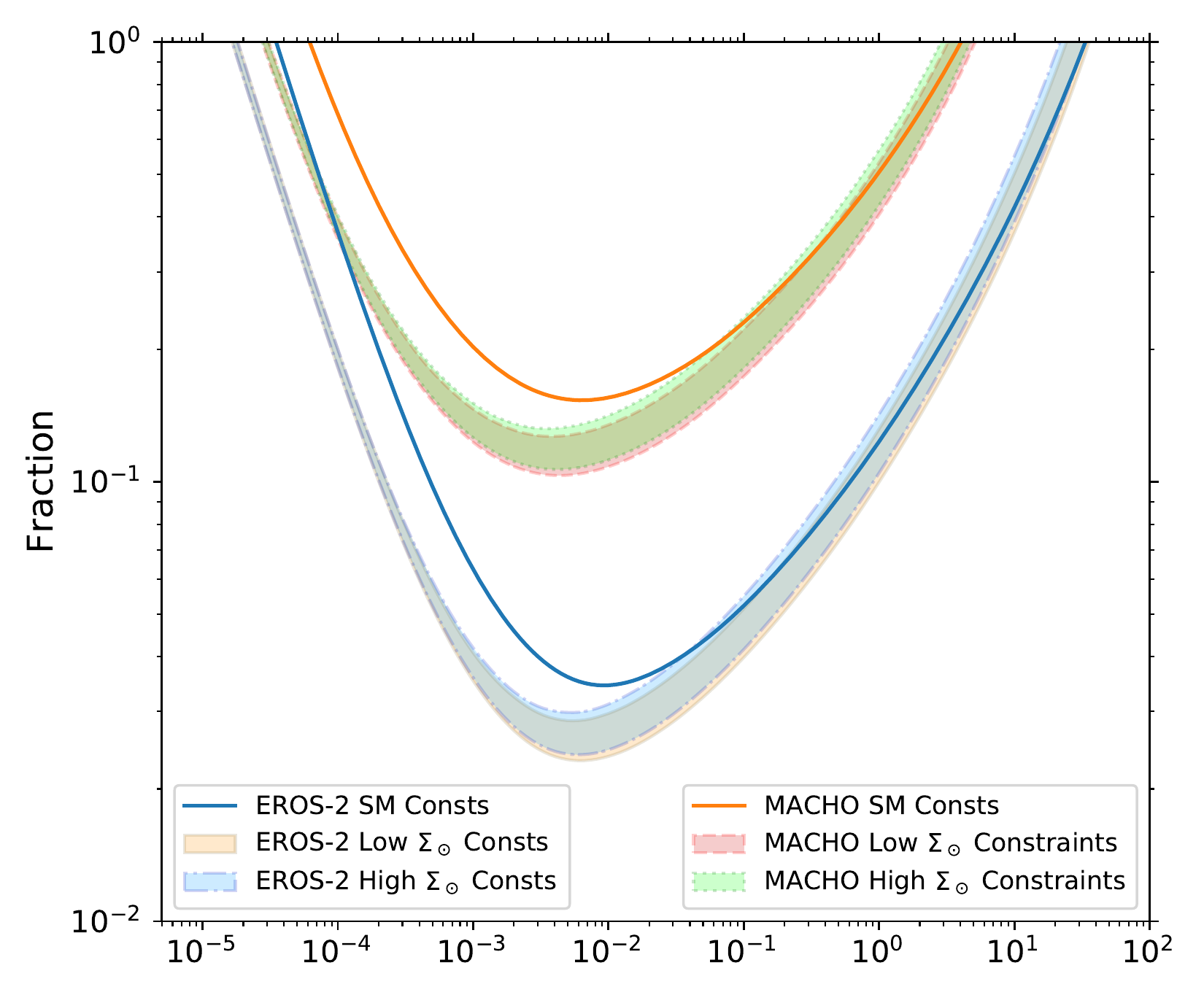}
\caption{The exclusion plots using the EROS-2 bright stars sample (bottom set of curves), in which no microlensing events were detected \citep{eros2007}, and the MACHO Collaboration 5.7 year results (top set of curves), with 10 events observed \citep{alcock2000,bennett2005}. The standard model constraints for each survey are plotted as solid lines, overlapping their respective updated constraint at high mass. The updated constraints for each survey are plotted as the shaded regions for both the low and high $\Sigma_\odot$ prior. The uncertainty from the MWRC fit (Figure \ref{fig:rot_plot}) is propagated to produce these shaded regions, but they also represent the uncertainty in the shape of the MW dark halo, as discussed in section \ref{sec:mwq}. The region enclosed within any of the U-shaped lines/contours is excluded at the 95\% confidence interval.}
\end{figure}

\begin{table*}
     \centering
     \caption{Our derivations of the standard model (old) limits compared with the new limits we obtain from fitting the MWRC. The limits presented here represent the strictest possible constraints on monochromatic masses based on the uncertainty propagated from the fits to the MWRC.}
    \label{tab:updated_conts}
    \begin{tabular}{ccc}
        \hline
	Survey & Standard Model Constraints & Our Constraints \\ 
	\hline
	MACHO & $6.18\times 10^{-5} \leq M/M_\odot \leq 4.04$  &    $3.07\times 10^{-5} \leq M/M_\odot \leq 2.87 $ \\ 
	EROS-2 & $3.49\times 10^{-5} \leq M/M_\odot \leq 33.5 $  &    $ 1.83\times 10^{-5} \leq M/M_\odot \leq 21.7$ \\
		\hline
    \end{tabular}
\end{table*}

The optical depth $\tau$ along the line of sight towards the LMC is plotted in Figure \ref{fig:tau_los}. The optical depth along the line of sight is higher for the newly derived models compared to the standard model. This is a direct result of a higher mass along the line of sight. The higher mass is partly due to the allowed oblateness of the dark halo. Despite having a falling rotation curve, it appears that the Milky Way dark halo contains more mass along the line of sight towards the LMC than predicted by the standard model.

\subsection{Monochromatic Mass Distribution}

The updated MACHO fraction constraints for both the EROS-2 and MACHO Collaboration surveys are presented in Figure \ref{fig:er_up}. Each of the shaded regions in Figure \ref{fig:er_up} represents the 2-$\sigma$ uncertainty from the MWRC fit in Figure \ref{fig:rot_plot}. Therefore, they represent the uncertainty in the 95\% confidence interval constraints by taking into account in the mass models that are used to build these constraints.

We show the individual microlensing constraints for the low and high $\Sigma_\odot$ constraints for both the EROS-2 and MACHO Collaboration surveys. In all cases, the constraints produced here are actually tighter at the lower-mass range by around half an order of magnitude. On the higher mass end, the constraints are mostly consistent with the standard model. When we consider the uncertainty in the mass models used to construct these, the constraints are marginally weaker for $M \gtrsim 0.1\,M_\odot$, but are stronger below this value.

The range that MACHOs are excluded from making the entirety of DM for both the standard model constraints and the updated power-law model constraints, are tabulated in Table \ref{tab:updated_conts}.

Despite being the most flexible of the three models tested in our analysis, the power-law model does have some limitations. The largest limitation is that it does not allow us to fully investigate different halo shapes, although it allows much more flexibility than the other models. The MWRC is best described by a falling rotation curve at large radii, and the power-law model is not able to produce a physical model for a wide range of halo shapes in this case. This means that we cannot encapsulate the full uncertainty that currently exists in the literature relating to the shape of the halo using the power-law model.

If future investigation reveals that the Milky Way dark halo does indeed have a strongly oblate, or even mildly prolate shape, new models for computing microlensing event rates will need to be derived. This is particularly important if future information on the MWRC is consistent with rotation curve produced by \cite{huang2016}, and the MWRC does indeed decrease at large galactocentric radii.

\subsection{Extended Mass Distribution and Clustered PBH Constraints}

\begin{figure*}
\centering
\label{fig:eros_ext_frac}
\includegraphics[width=1.\textwidth]{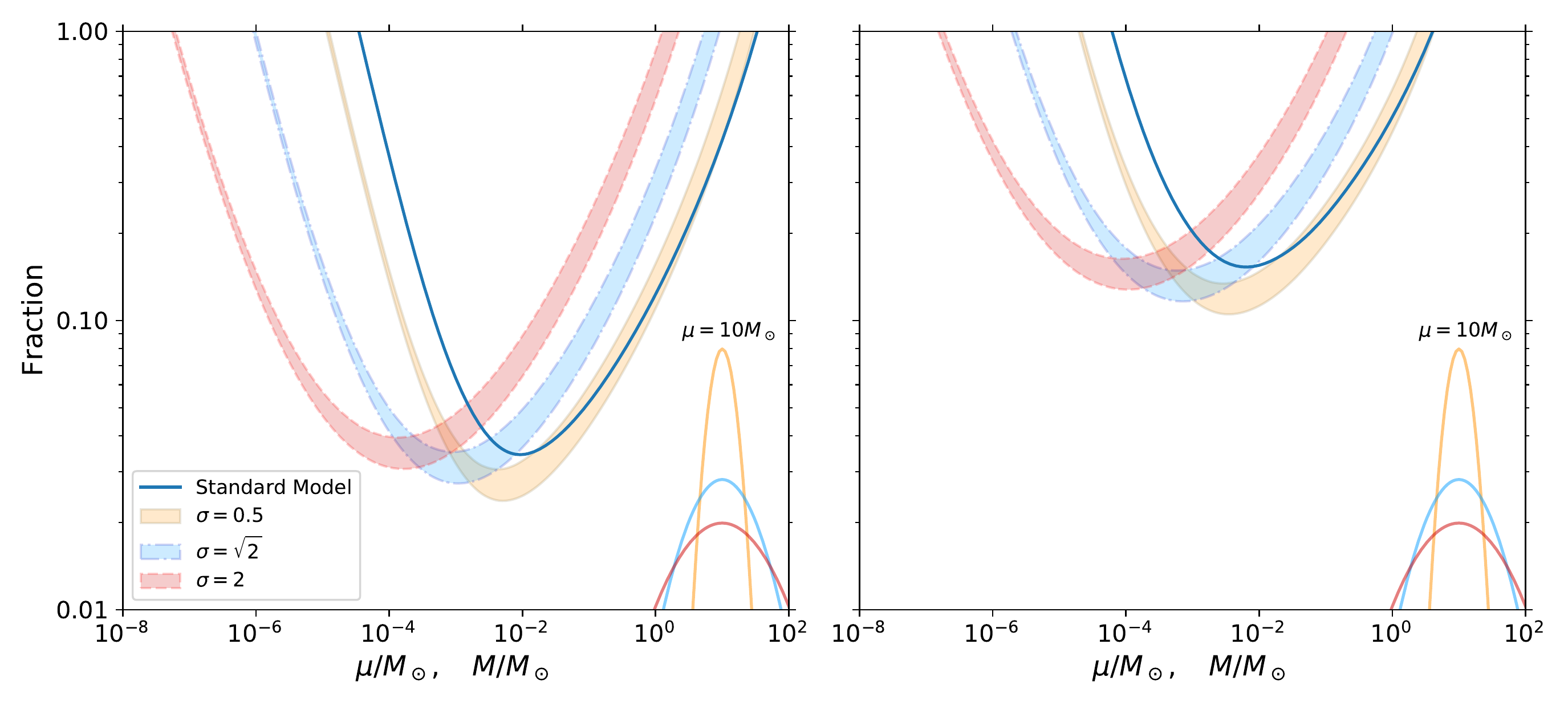}
\caption{The exclusion plots for the EROS-2 (left) and MACHO (right) surveys when considering an extended mass function. The shaded regions plotted represent a combination of the low and high $\sigma_\odot$ priors in Figure \ref{fig:er_up}. The $x$-axis represents the peak mass $\mu$ for the extended mass distribution constraints. The standard model constraints for the monochromatic (delta function) masses are plotted as a blue solid line, and only slightly deviate from the extended mass distribution with a small dispersion $\sigma=0.5$. It can be noted that the constraints from both surveys weaken substantially for broader mass distributions for the updated constraints. On the right hand side of each plot we have plotted the mass distributions for $\mu = 10\,M_\odot$ and $\sigma=[0.5, \sqrt{2}, 2]$, but have scaled them by $e^{\frac{1}{2}\sigma^2}$ so that they are visible on the $y$-scale of this figure.}
\end{figure*}

\begin{figure*}
\centering 
\label{fig:eros_clust_frac}
\includegraphics[width=1.\textwidth]{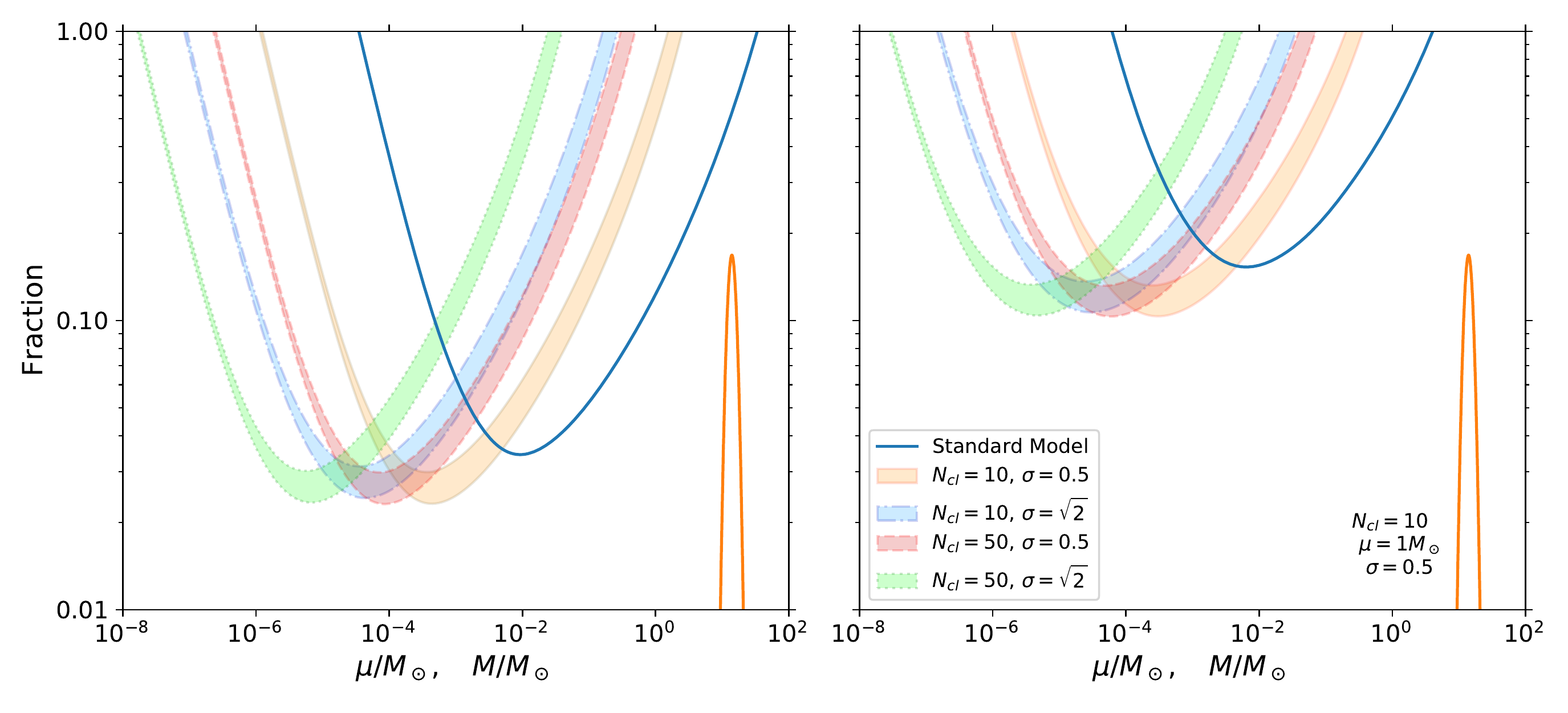}
\caption{The exclusion plots for the EROS-2 (left) and MACHO (right) surveys when considering clustered PBHs. The $x$-axis represents the peak mass $\mu$ for the extended mass distribution that makes the cluster of PBH (see equation \ref{eq:newms}). The standard model constraints for the monochromatic (delta function) masses is plotted in blue. A larger cluster size $N_{\textrm{cl}}$ is more effective at shifting the microlensing constraints towards lower masses due to the higher combined mass of the cluster. A clustered PBH distribution is plotted to the right with parameters $N_{\rm cl}=10$, $\mu = 1\, M_\odot$, $\sigma = 0.5$.}
\end{figure*}

Here we consider the effect of an extended mass function and clustered PBHs. The updated MACHO constraints from the EROS-2 (left) and MACHO (right) surveys are presented in Figure \ref{fig:eros_ext_frac}. Each of the shaded regions represent the 2-$\sigma$ uncertainty region of the MWRC fit using the power-law model, for a given extended mass distribution with some dispersion $\sigma$. Starting with $\sigma=0.5$, we can see that when the dispersion is small, the constraints in Figure \ref{fig:eros_ext_frac} do not strongly deviate away from when we considered the monochromatic PBH masses (comparing the tan shaded region in Figure \ref{fig:eros_ext_frac} to the shaded region in Figure \ref{fig:er_up}). For small dispersions, the lognormal distribution is well approximated by a delta-function. The scenario changes when the dispersion increases. The general trend is that the constraints shift towards the lower masses since when the dispersion is increasing, some PBH end up at higher masses where they essentially become ``undetectable". That is, some PBH end up at such high masses that they are not efficiently detectable unless the observational baseline of the microlensing survey increases. Since a small proportion of the detectable population is removed from this, the constraints slightly begin to weaken (i.e. the tan shaded region, which represents a narrow PBH distribution, is more constraining than the red shaded region, which represents a broad PBH distribution).

A much more drastic result is seen with the clustered PBHs in Figs.~\ref{fig:eros_ext_mass} and \ref{fig:eros_clust_frac}. Here, even a small dispersion and small cluster size with $N_{\rm cl}=10$ PBHs can shift the constraints substantially. With larger cluster sizes ($N_{\rm cl}=50$), and a broader mass distribution ($\sigma=\sqrt{2}$), the constraints shift to even lower masses. As discussed above, in section~\ref{sec:WMD}, the clustering of PBH produces a concentration of mass on smaller-sized objects spread over a larger volume, substantially diluting the microlensing constraints from $C(M)$ to $C(N_{\rm cl}\langle M\rangle)$. This is the reason for the significant shift of the constraints to lower masses in this case, as seen in Fig.~\ref{fig:eros_clust_frac}.

The constraints presented here are made with the assumption that the cluster size is much smaller than the Einstein radius, meaning that the microlensing signal they produce is similar to the regular point source/lens single as in equation (\ref{eq:amp}), often referred to as a Paczynski light-curve. If this assumption breaks down, the microlensing light-curves can start to behave in a complicated fashion that is not encapsulated by the simple  point source/lens case. For example, a cluster that is spatially large can cause caustics, bright and sudden spikes in the microlensing light-curve. If a cluster has many objects, there may be many caustics occurring over short time intervals, and probably would not be picked up by a search algorithm that is optimised to recover the simple Paczynski light-curve. The efficiency functions from both the MACHO and EROS-2 collaborations are determined using the point source/lens assumption, and so any constraints derived with these efficiency functions are subject to this assumption.

\section{Summary}\label{sec:sum}
In this paper we derived constraints on the Milky Way rotation curve (MWRC) using the recent compilation by \cite{huang2016}. We then used these constraints to update the MACHO fraction of the dark halo, for monochromatic and extended mass distributions. We have found that the standard model that has previously been used to derive MACHO constraints is no longer an accurate description of the MWRC. We propagate the uncertainties associated with the MWRC fits when constraining the MACHO fraction within the dark halo. This produces updated constraints on the MACHO fraction in the halo, which is more reflective of the current uncertainties associated with the size and shape of the Milky Way galaxy (both luminous and dark).

We update the microlensing constraints from the EROS-2 LMC bright stars sample \citep{eros2007} and the MACHO Collaboration 5.7 year results \citep{alcock2000,bennett2005}, and summarise our updated constraints in table \ref{tab:updated_conts}. The constraints at the higher mass end are somewhat weakened when the uncertainty in the Milky Way mass models are taken into account. At lower masses the constraints are actually tighter.

When considering an extended mass distribution, the constraints from microlensing weaken substantially with a broad distribution. Taking into account the possibility of clustered PBHs loosens the constraints further, even for small clusters with only $N_{\rm cl} = 10$ PBHs. It is expected that any astrophysical population will have some dispersion in their intrinsic properties. Therefore, when considering microlensing constraints, both the possible spatial distribution and mass distribution in the lens population should be taken into account.

Our analysis highlights the importance of mass modelling when deriving microlensing constraints. In the case of constraining primordial black holes as dark matter, we have demonstrated that the uncertainties relating to the size and shape of the Milky Way dark halo should be taken into account to produce more accurate constraints. This will be especially important when one wants to constrain parameters on the lens population. Not properly accounting for known uncertainties will bias the result. Even if the objective of microlensing searches is not to uncover the nature of dark matter, understanding the physical and dynamical structure of the Milky Way is important for obtaining an accurate understanding of a lensing population.

One source of uncertainty that we have not been able to properly account for is the shape of the Milky Way halo. We have made an attempt, within the framework of the power-law model, to account for this. However, since the MWRC data prefers models that produce a falling rotation velocity with distance, the flexibility in the shape of the power-law model is restricted (see Figure \ref{fig:df_plot}).

If future exploration of the Milky Way dark halo reveals a strong deviation away from spherical symmetry (particularly in the prolate case), other models that offer more flexibility than that of the power-law model should be adopted for determining microlensing event rates.

\section*{Acknowledgements}
We thank our colleagues in the Dark Energy Survey for interesting discussions that sparked the line of enquiry in this paper, particularly James Annis, Marcelle Soares-Santos, Eric H Neilsen Jr., and Alex Drlica-Wagner. We also thank the referee, Michael Hawkins, for their useful comments that helped to improve the quality of this manuscript.

This paper makes use of the Python package \texttt{ChainConsumer} for plotting MCMC chains \citep{hinton2017}. JC would like to acknowledge support by an Australian Government Research Training Program (RTP) Scholarship for this research. Parts of this research were conducted by the Australian Research Council Centre of Excellence for All-sky Astrophysics (CAASTRO), through project number CE110001020. Part of this work is supported by the Spanish Research
Project FPA2015-68048-C3-3-P [MINECO-FEDER] and
the Centro de Excelencia Severo Ochoa Program SEV-
2012-0597. JGB thanks the Theory Department at
CERN for their hospitality during a Sabbatical year at
CERN. He also acknowledges support from the Salvador
de Madariaga Program Ref. PRX17/00056.





\bibliographystyle{mnras}
\bibliography{paper}


\appendix
\section{Results of Rotation Curve Fits}
\begin{figure*}
\centering
\label{fig:ISO}
\includegraphics[width=1.\textwidth]{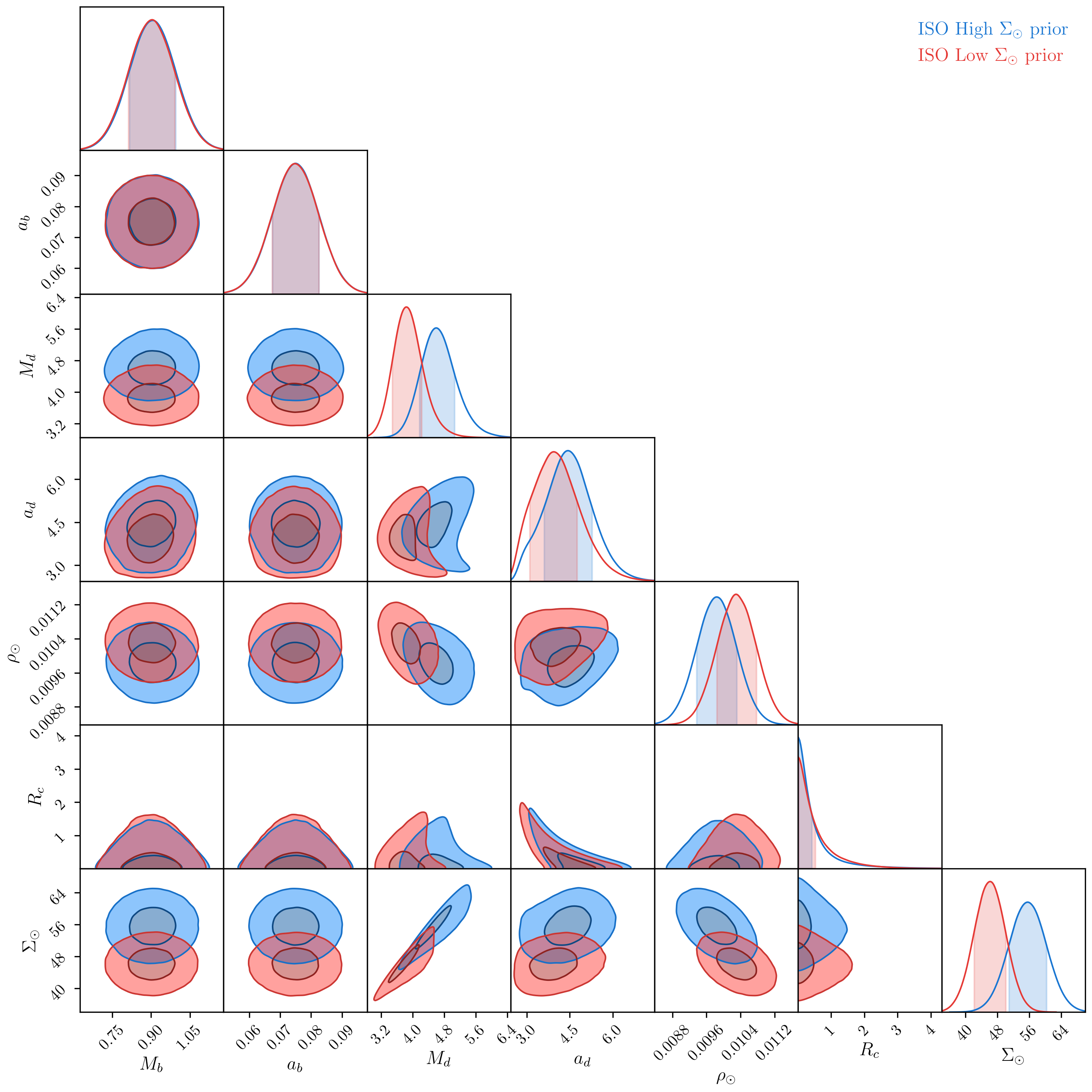}
\caption{The contour plot for the constraints on the semi-isothermal dark halo model. The blue and red regions are the fits for the high and low $\Sigma_\odot$ priors respectively. }
\end{figure*}

\begin{figure*}
\centering
\label{fig:NFW}
\includegraphics[width=1.\textwidth]{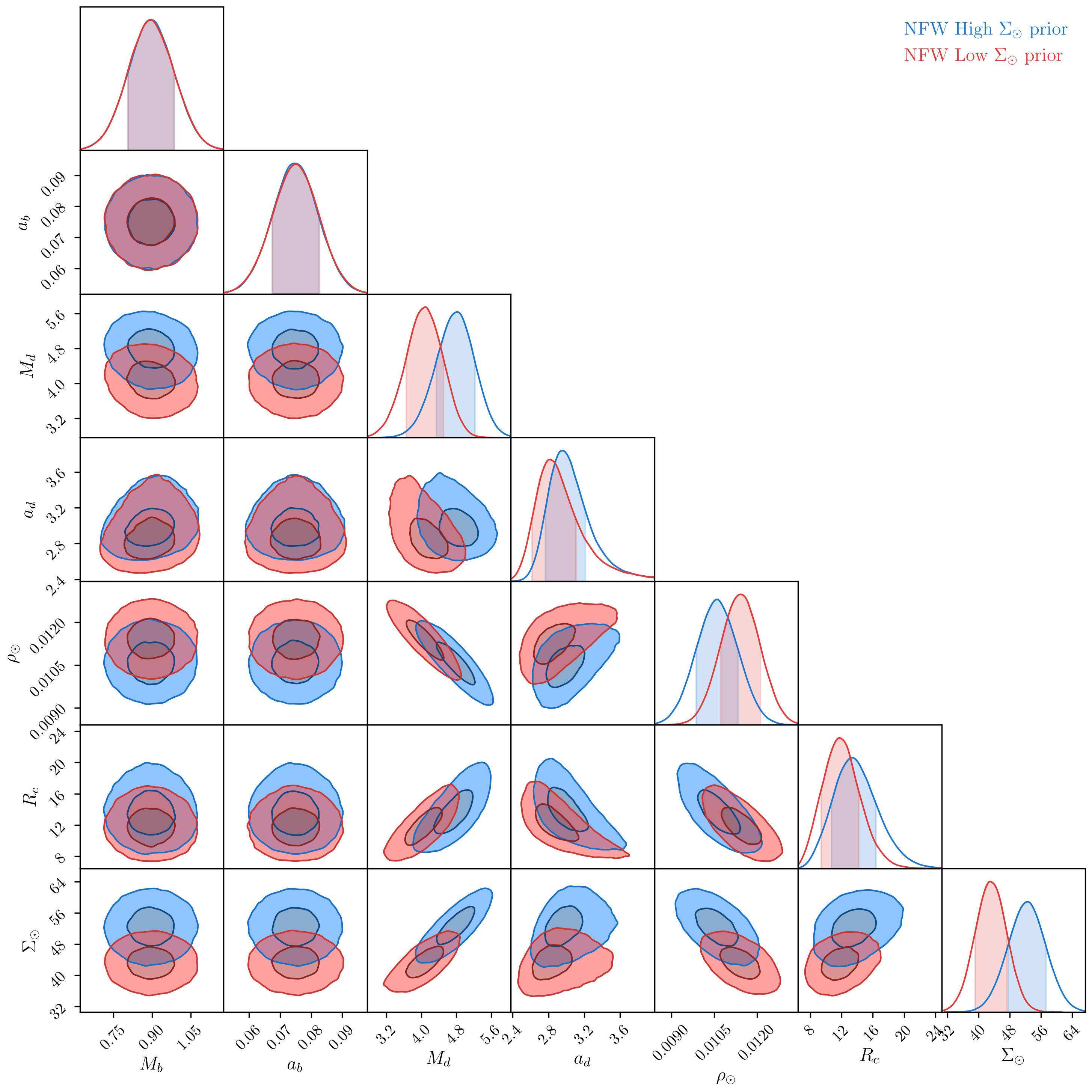}
\caption{As with Figure \ref{fig:ISO}, but for the NFW dark halo profile. }
\end{figure*}

\begin{figure*}
\centering
\label{fig:PL}
\includegraphics[width=1.\textwidth]{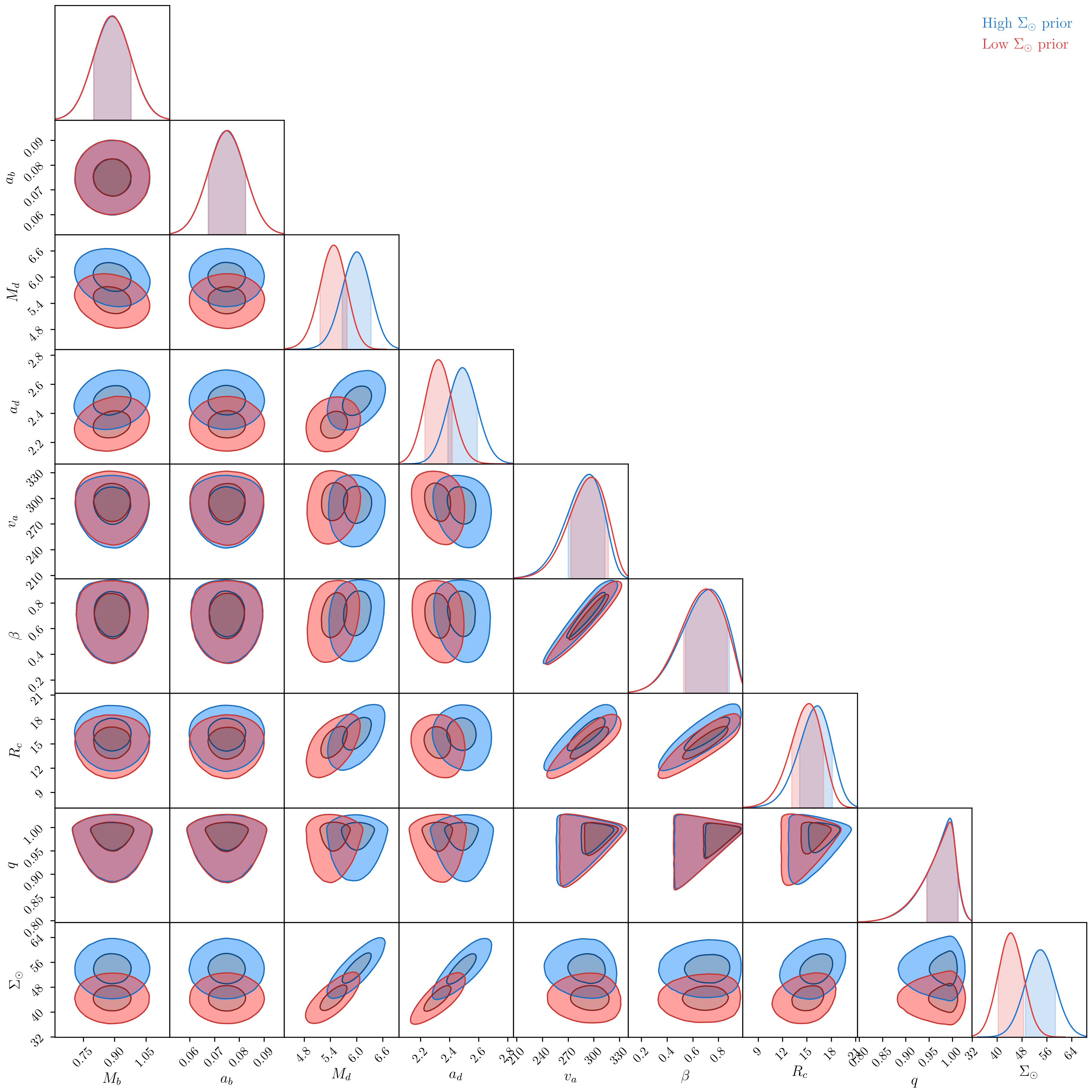}
\caption{The contour plot for the constraints on the power-law dark halo model. The rather triangular contours for the $q$ parameter are a result of the hard-wall prior forcing each step in the MCMC chain to only sample models that are self-consistent (see section \ref{sec:plm}). A large shift in the disk properties ($a_d,\, M_d$), results in only a small shift in the parameters for the dark halo fit ($v_a,\, R_d,\, \beta$).}
\end{figure*}


\label{lastpage}
\end{document}